\documentclass[twocolumn,twocolappendix]{aastex631}

\usepackage{amsmath}
\usepackage{amssymb}
\usepackage{bm}

\newcommand{\diff}{\ensuremath{\mathrm{d}}}
\newcommand{\e}{\mathrm{e}}

\newcommand{\lambertW}{\mathrm{W}_\mathrm{L}}
\newcommand{\arsinh}{\mathrm{arsinh}}
\newcommand{\artanh}{\mathrm{artanh}}

\newcommand{\acoll}{a_\mathrm{coll}}
\newcommand{\tcoll}{t_\mathrm{coll}}

\newcommand{\kcut}{k_\mathrm{cut}}

\newcommand{\rhos}{\rho_\mathrm{s}}
\newcommand{\rs}{r_\mathrm{s}}

\newcommand{\Achar}{\tilde{A}}
\newcommand{\mchar}{\tilde{m}}

\newcommand{\Msol}{\mathrm{M}_\odot}
\newcommand{\kpc}{\mathrm{kpc}}

\shorttitle{The cusp-halo relation}
\shortauthors{M. S. Delos}
\graphicspath{{./}{figures/}}

\begin{document}

\title{The cusp-halo relation}

\email{mdelos@carnegiescience.edu}

\author[0000-0003-3808-5321]{M. Sten Delos}
\affiliation{Carnegie Observatories, 813 Santa Barbara Street, Pasadena, CA 91101, USA}

\begin{abstract}
	
Simulations have established that each halo of collisionless dark matter is expected to contain a $\rho = A r^{-1.5}$ density cusp at its center.
This \textit{prompt cusp} is a relic of the halo's earliest moments and has a mass comparable to the cutoff scale in the spectrum of initial density perturbations.
In this work, we provide a framework to predict for each halo the coefficient $A$ of its central cusp.
We also present a ``cusp-NFW'' functional form that accurately describes the density profile of a halo with a prompt cusp at its center.
Accurate characterization of each halo's central cusp is of particular importance in the study of warm dark matter models, for which the spectral cutoff is on an astrophysically relevant mass scale.
To facilitate easy incorporation of prompt cusps into any halo modeling approach, we provide a code package that implements the cusp-halo relation and the cusp-NFW density profile.

\end{abstract}


\section{Introduction} \label{sec:intro}

\begin{figure}
	\centering
	\includegraphics[width=\columnwidth]{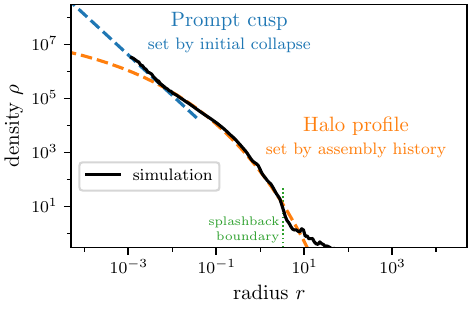}
	\caption{Radial structure of a dark matter halo. The black curve shows the spherically averaged density profile of the simulated halo W1 from \citet{2023MNRAS.518.3509D} in arbitrary units. The dashed blue line marks the prompt cusp predicted from the initial conditions, while the dashed orange curve is a fitted Einasto profile.}
	\label{fig:profile-diagram}
\end{figure}

Simulations have revealed that halos of collisionless dark matter contain historical records in their radial structures.
Figure~\ref{fig:profile-diagram} shows an example of this structure.
An outermost splashback boundary connects to the accretion rate over the very recent past, corresponding to about one crossing time \citep{2017ApJ...843..140D,2023MNRAS.521.5570S}.
Inside the halo, the radial density profile is tightly linked to the history of the halo's assembly over cosmic time \citep{2013MNRAS.432.1103L,2016MNRAS.460.1214L}.
Density profiles tend to exhibit a steeply falling density at large radii, which connects to a period of gradual assembly at late times, and transition towards a more shallowly varying density at small radii, which is linked to an early phase of rapid accretion \citep{2002ApJ...568...52W,2006MNRAS.368.1931L}.
Finally, at the center of each halo lies a relic of its initial formation event: the prompt $\rho\propto r^{-1.5}$ density cusp \citep{2019PhRvD.100b3523D,2023MNRAS.518.3509D,2023JCAP...10..008D,2024MNRAS.52710802O}.

Prompt cusps are a signature of dark matter microphysics.
Although dark matter's particle identity remains unclear, its properties would manifest themselves in the spatial distribution of the dark matter prior to structure formation. In particular, the mass and interaction properties of the dark matter would give rise to a fundamental lower limit or ``cutoff'' to the length scales of density variations, as smaller-scale perturbations are erased by thermalization processes \citep[e.g.][]{1983ApJ...274..443B,2001ApJ...556...93B,2009NJPh...11j5027B,2016PhRvD..93l3527C}.
Prompt cusps form at this cutoff scale from the collapse of local maxima in the initial density field \citep{2019PhRvD.100b3523D,2022MNRAS.517L..46W,2023MNRAS.518.3509D,2023JCAP...10..008D,2023PDU....4101259D,2024MNRAS.52710802O}. Over time, halos grow around these cusps as they continue to accrete material \citep{2023MNRAS.518.3509D}.

The parameters of each prompt cusp are tightly linked to the local features of the initial density maximum from which it formed \citep{2019PhRvD.100b3523D,2023MNRAS.518.3509D,2024MNRAS.52710802O}. Consequently, the overall cosmological distribution of prompt cusps may be quantified straightforwardly using the statistics of peaks in random fields \citep[e.g.][]{1986ApJ...304...15B}.
Previous studies used this approach to evaluate the contribution of the cusps to dark matter annihilation, both for minimal concordance cosmologies \citep{2023JCAP...10..008D,2023MNRAS.523.1067S,2024PhRvD.109b3532D,2024PhRvD.109h3512D,2025arXiv250114865C} and for nonstandard early-Universe thermal histories \citep{2019PhRvD.100l3546D,2023PhRvD.108b3528D,2024arXiv240318893G,2025arXiv250208719B}.

It is a more subtle matter to predict the central cusp of a given dark matter halo. That is the subject of the present work.
Understanding each halo's central cusp is of particular importance in the context of warm dark matter models, for which the small-scale cutoff is of similar order to the mass scales of the smallest galaxies.
Numerous recent works have used the abundance and properties of low-mass subhalos of the Milky Way and other systems to place limits on these models
\citep{2020MNRAS.491.6077G,2020MNRAS.492.3047H,2021JCAP...10..043B,2021MNRAS.506.5848E,2021ApJ...917....7N,2021PhRvL.126i1101N,2021JCAP...08..062N,2022PhRvL.129s1301Z,2024MNRAS.535.1652K,2024arXiv241003635N}.
The signature of warm dark matter that these works considered is a suppression in the abundance and internal density of halos with masses below or close to the cutoff scale, as reported in earlier simulation studies \citep[e.g.][]{2013MNRAS.433.1573S,2013MNRAS.434.3337A,2014MNRAS.439..300L,2016MNRAS.455..318B,2018JCAP...04..010L,2019ApJ...874..101S,2020ApJ...897..147L,2021MNRAS.506..128B,2022MNRAS.509.1703S}.
However, as \citet{2023MNRAS.522L..78D} noted, the massive prompt cusps that arise in warm dark matter scenarios would produce an opposing effect: an enhancement to the internal density of small subhalos.
This effect must be accounted for to place accurate observational constraints on these models, and it could also be used as an independent test of warm dark matter, as suggested by \citet{2023MNRAS.522L..78D}.
The purpose of the present work is to provide a framework for easily incorporating the central prompt cusp into a description of dark matter halos.

\begin{figure}
	\centering
	\includegraphics[width=\columnwidth,trim={0 11cm 20.5cm 0},clip]{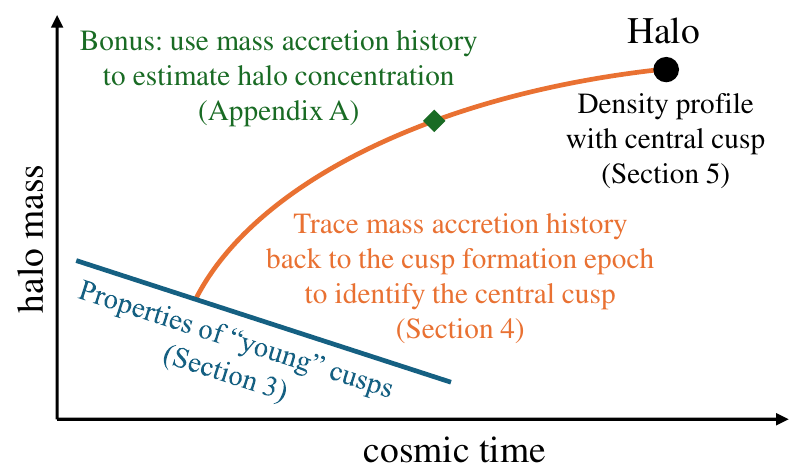}
	\caption{Outline of how we approach the cusp-halo relation. Starting with a halo of a given mass at some late time (black), we can trace its growth history backward in time (orange) until we meet the mass of a newly formed prompt cusp (blue), which depends on the epoch. The halo may then be assigned a central cusp corresponding to the intersection point. We can also obtain from this process an estimate of the halo's density profile (parameterized by the concentration), since that is set by the mass accretion history.}
	\label{fig:diagram}
\end{figure}

We build the \textit{cusp-halo relation} using a strategy outlined in figure~\ref{fig:diagram}.
At each time, one can characterize the distribution of newly formed, ``young'' prompt cusps (blue). This distribution follows approximately, albeit not exactly, from the statistics of peaks in the initial density field. For a given halo at a much later time (black), one may now trace its mass accretion history backwards through time (orange) until it meets the distribution of young cusps. The intersection point yields an estimate of the halo's central cusp.

We also present a ``cusp-NFW'' functional form that accurately describes the combined density profile of a halo with a prompt cusp at the center.
Density profiles of dark matter halos are usually modeled with the Navarro-Frenk-White (NFW) form \citep{1996ApJ...462..563N,1997ApJ...490..493N},
\begin{equation}\label{NFW}
	\rho(r)
	=
	x^{-1}(1+x)^{-2}\rhos,
	\qquad x=r/\rs,
\end{equation}
or the \citet{1965TrAlm...5...87E} form \citep{2004MNRAS.349.1039N},
\begin{equation}\label{einasto}
	\rho(r)
	=
	\e^{-\frac{2}{\alpha_\mathrm{E}}(x^{\alpha_\mathrm{E}}-1)}\rhos,
	\quad x=r/\rs, \quad \alpha_\mathrm{E}\simeq 0.17,
\end{equation}
where $\rhos$ and $\rs$ are fitting parameters that are different for each halo. We find that a simple modification to the NFW form, given by
\begin{align}\label{cuspNFW0}
	\rho(r) = \frac{\sqrt{y^2+x}}{x^{1.5}(1+x)^2}\rhos, \quad
	x = r/\rs, \quad 0\leq y\leq1,
\end{align}
provides an accurate fit to numerical $N$-body simulations. Here $y$ is a new halo-dependent parameter that quantifies the relative density and size of the central cusp in relation to the density and size of the halo.

Figure~\ref{fig:diagram} summarizes how the main parts of this work are organized.
In section~\ref{sec:setup}, we detail the cosmological simulations and the analysis procedures that we use for this work.
In section~\ref{sec:cusps}, we characterize the properties of newly formed cusps in each epoch.
In section~\ref{sec:cusphalo}, we develop the cusp-halo relation by using halos' mass accretion histories to match them with prompt cusps formed in earlier epochs.
In section~\ref{sec:profile}, we present the cusp-NFW density profile and show that it accurately describes halos with central prompt cusps.
We summarize the results and conclude in section~\ref{sec:conclusion}.
Appendix~\ref{sec:concentration} explores the halo mass accretion histories in more detail, showing that they can be used to make a simple but reasonably accurate estimate of halo concentration parameters, which characterize their broader density profiles.
Finally, we show in appendix~\ref{sec:stability} that the cusp-NFW profile describes a system that is dynamically stable.

For ease of use, we provide a \textsc{Python} package at \url{https://github.com/delos/cusp-halo-relation} \citep{delos_2025_17064050} that implements the cusp-halo relation, halo concentration estimates, and calculations related to the cusp-NFW profile.

\section{Technical setup}\label{sec:setup}

We first detail the simulations and analysis procedures used in this work. We rely on the results of \citet{2019PhRvD.100b3523D}, \citet{2023MNRAS.518.3509D}, and \citet{2024MNRAS.52710802O}, who showed in high-resolution simulations that the properties of prompt cusps are tightly linked to those of the initial density peaks from which they form.\footnote{Central cusps with slopes approaching $\rho\propto r^{-1.5}$, or similar values, were reported in many earlier works before their connection to initial density peaks was clear. These earlier works include \citet{2005Natur.433..389D}, \citet{2010ApJ...723L.195I}, \citet{2013JCAP...04..009A}, \citet{2014ApJ...788...27I}, \citet{2016MNRAS.461.3385O}, \citet{2017MNRAS.471.4687A}, and \citet{2020MNRAS.492.3662I}, who studied the smallest halos of cold dark matter; \citet{2015MNRAS.450.2172P}, who studied halos of warm dark matter; \citet{2018PhRvD..97d1303D,2018PhRvD..98f3527D}, who studied early-forming halos in nonstandard cosmologies; and \citet{2018MNRAS.473.4339O} and \citet{2021A&A...647A..66C}, who explored general aspects of gravitational structure formation.} Consequently, we do not need to resolve in detail the structure of the prompt cusp residing at the center of each halo. Instead, we focus on tracking from which initial density peak each halo originated.

\subsection{Simulations}\label{sec:simulations}

\begin{figure}
	\centering
	\includegraphics[width=\columnwidth]{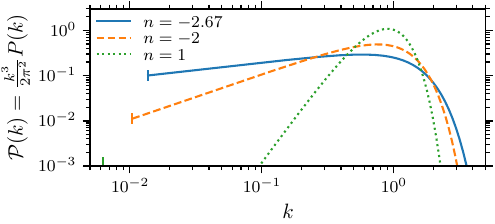}
	\caption{Linear-theory matter power spectra for the simulations used in this work, shown in dimensionless form at the time when $\sigma_0=1$. The vertical markers indicate the smallest wavenumber represented in the simulation volume, $k=2\pi/L_\mathrm{box}$. Units are given in table~\ref{tab:units}.}
	\label{fig:simpower}
\end{figure}

We use the simulations from \citet{2023MNRAS.518.3509D}, which represent idealized dark-matter-dominated universes with the matter power spectra
\begin{align}
	P(k) \propto k^n \e^{-(k/\kcut)^2}
\end{align}
for $n=-2.67$, $n=-2$, and $n=1$.
These power spectra correspond to density fields that are gaussian-smoothed on a length scale $\kcut^{-1}$. They are shown in figure~\ref{fig:simpower}. Although they are idealized, these spectra are representative of a wide range of cosmologies on scales close to the cutoff in the power spectrum caused by dark matter free streaming. The $n=-2.67$ and $n=-2$ power spectra are similar to traditional models of cold dark matter \citep[e.g.][]{2004MNRAS.353L..23G,2005JCAP...08..003G,2005PhRvD..71j3520L,2006PhRvL..97c1301P,2006PhRvD..74f3509B,2007JCAP...04..016B} and warm dark matter \citep[e.g.][]{2001ApJ...556...93B,2002MNRAS.333..544H,2005PhRvD..71f3534V,2023PhRvD.108d3520V} close to the free-streaming scale. The $n=1$ power spectrum lies at an opposite extreme and is the spectrum expected at small scales due to an early matter-dominated era \citep[e.g.][]{2011PhRvD..84h3503E,2015PhRvD..92j3505E,2021PhRvD.103j3508E,2022JCAP...01..017E,2023JCAP...01..004G,2024JCAP...04..015G} or certain models of cosmic inflation \citep{2019JCAP...06..028B,2024JCAP...05..022C,2024arXiv241022154C}.
It is also close to the ($n=0$) spectrum that would result from Poisson noise in the dark matter distribution \citep[e.g.][]{2003ApJ...594L..71A,2007PhRvD..75d3511Z,2025PhRvD.111e5019K,2025arXiv250320881A}.

The power spectrum supplies a system of natural units. Following the notation of \citet{1986ApJ...304...15B}, we define
\begin{align}\label{sigmaj}
	\sigma_j^2 = \int_0^\infty \frac{\diff k}{k} \mathcal{P}(k) \, k^{2j},
\end{align}
where $\mathcal{P}(k)=\frac{k^3}{2\pi^2}P(k)$ is the dimensionless form of the matter power spectrum.
We will find it most natural to use physical (not comoving) units everywhere in this work, so the wavenumber $k$ in equation~(\ref{sigmaj}) is the physical (not comoving) wavenumber. This means that, for a matter dominated universe, $\sigma_j$ evolves over time as $\sigma_j\propto a^{1-j}$, where $a$ is the cosmic expansion factor.
Note that $\sigma_j$ has dimensions of $(\mathrm{length})^{-j}$.

The rms density contrast $\sigma_0$ is a natural time parameter which grows proportionally with the expansion factor $a$.
Length units can be constructed out of the $\sigma_j$ for $j>0$, but we will see that prompt cusps are most naturally connected to $\sigma_2$. Therefore, $(\sigma_0/\sigma_2)^{1/2}$ is the appropriate comoving length unit, while $(\sigma_0\sigma_2)^{-1/2}$ is the corresponding physical length unit. By combining these length units with $\bar\rho$, the mean density of the universe, we can also obtain mass units. Table~\ref{tab:units} summarizes the system of natural units. For the sake of brevity, we will use these units implicitly in many of the figures in this work. However, they will always be written out explicitly in the body of the article.
We will find that not only are the properties of prompt cusps nearly universal (across different cosmologies) in these units, but so are the abundance of cusps and the growth histories and merger histories of dark matter halos.

The simulations were executed using the \textsc{gadget-4} simulation code \citep{2021MNRAS.506.2871S} in periodic boxes of comoving length $L_\mathrm{box}=643\kcut^{-1}$. Note that $\kcut\simeq 1.41(\sigma_0/\sigma_2)^{-1/2}$, $1.07(\sigma_0/\sigma_2)^{-1/2}$, and $0.64(\sigma_0/\sigma_2)^{-1/2}$ for $n=-2.67$, $n=-2$, and $n=1$, respectively.
When evaluating the $\sigma_j$ in equation~(\ref{sigmaj}), we integrate the power spectra only over $k>2\pi/L_\mathrm{box}$, where $L_\mathrm{box}$ is the simulation box size, since only these modes contribute to the initial conditions in the simulations.\footnote{The units used here are similar to those in \citet{2023MNRAS.518.3509D}, but they differ in that \citet{2023MNRAS.518.3509D} used $\sigma_0/\sigma_1$ as the comoving length unit instead of $\sqrt{\sigma_0/\sigma_2}$ and did not restrict the integration range to $k>2\pi/L_\mathrm{box}$ when evaluating the $\sigma_j$.}

\begin{table}
	\centering
	\caption{The natural units that we employ in many of the figures. We only use physical units, but the table includes comoving length and density units for completeness. Here $\bar\rho$ is the cosmological mean matter density, and the $\sigma_j$ are defined in equation~(\ref{sigmaj}). We will also use $\sigma_0\propto a$ as a time parameter.}
	\label{tab:units}
	\begin{tabular}{ccc}
		\hline
		quantity & unit (comoving) & unit (physical) \\
		\hline
		length & $(\sigma_0/\sigma_2)^{1/2}$ & $(\sigma_0\sigma_2)^{-1/2}$ \\
		density & $\bar\rho$ & $\bar\rho\sigma_0^3$ \\
		mass & - & $\bar\rho(\sigma_0/\sigma_2)^{3/2}$ \\
		cusp $A$ & - & $\bar\rho\sigma_0^{9/4}\sigma_2^{-3/4}$ \\
		\hline
	\end{tabular}
\end{table}

For most of the results of this work, we employ the ``primary'' $1024^3$-particle simulations from \citet{2023MNRAS.518.3509D}, as opposed to the high-resolution subvolume resimulations. For these primary simulations, interparticle forces were softened at the length scale $\epsilon=0.03L_\mathrm{box}/1024$ (0.03 times the initial interparticle spacing). Note that for the force-softening kernel used by \textsc{gadget-4}, forces are sub-Newtonian below a distance of $2.8\epsilon$. Table~\ref{tab:sim} summarizes the main numerical resolution parameters for these simulations.

\begin{table}
	\centering
	\caption{Resolution parameters for the primary simulations of \citet{2023MNRAS.518.3509D} in the natural units of table~\ref{tab:units}. $L_\mathrm{box}$ is the simulation box size, $\epsilon$ is the softening length (so forces are sub-Newtonian below $2.8\epsilon$), and $m_\mathrm{p}$ is the particle mass.}
	\label{tab:sim}
	\begin{tabular}{lrrr}
		\hline
		simulation & $L_\mathrm{box}$ & $2.8\epsilon$ & $m_\mathrm{p}$ \\
		\hline
			$n=-2.67$ & $455$ & $0.037$ & $0.088$ \\
			$n=-2$ & $600$ & $0.049$ & $0.202$ \\
			$n=1$ & $1007$ & $0.083$ & $0.951$ \\
		\hline
	\end{tabular}
\end{table}

\begin{figure}
	\centering
	\includegraphics[width=\columnwidth]{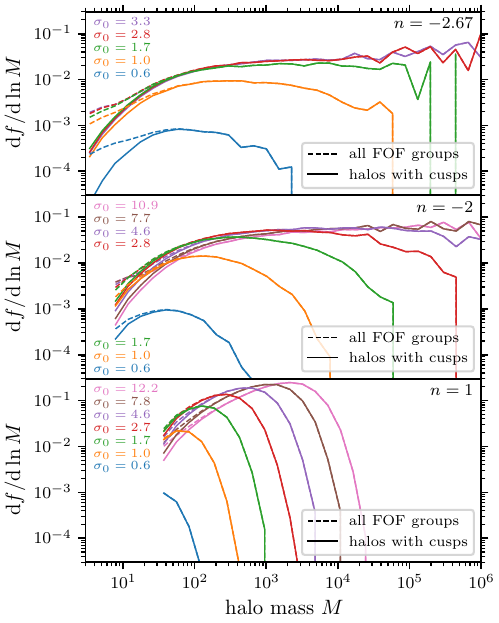}
	\caption{Halo mass functions in the three simulations (different panels), expressed as the differential mass fraction $\diff f/\diff\ln M$ in halos of mass $M$. We show a range of times, which are parameterized by $\sigma_0$, and we use the $M_{200}$ mass definition; see table~\ref{tab:units} for units. Dashed curves show the mass function of all FOF groups, which include spurious halos formed by artificial fragmentation. The solid curves are restricted to groups to which we can assign central prompt cusps; this procedure should exclude numerical artifacts. Mass functions are binned in intervals of $\Delta\ln M\simeq 0.4$, and we only consider masses $M>32m_\mathrm{p}$, where $m_\mathrm{p}$ is the simulation particle mass.}
	\label{fig:massfunction}
\end{figure}

Halos in the simulations were identified using the friends-of-friends (FOF) group finder in \textsc{gadget-4} with a comoving linking length of 0.15 times the initial interparticle spacing.\footnote{FOF linking lengths close to 0.15 have been found to be appropriate during the early stages of structure formation to avoid ``overlinking'' halos that have not (yet) merged \citep{2008MNRAS.385.2025C,2013MNRAS.433.1230W}. However, we verified that the main results of this work are unchanged when a more common linking length of 0.2 is adopted instead.}
These FOF groups represent systems that are enclosed by a surface of uniform density $\rho\simeq 140\bar\rho$ \citep[e.g.][]{1994MNRAS.271..676L}, so they can be interpreted as (field) halos. The dashed lines in figure~\ref{fig:massfunction} show the halo mass functions in each simulation using these FOF groups. Specifically, we show $\diff f/\diff\ln M$, the fraction of mass in halos of mass $M$ per logarithmic interval in $M$; the corresponding halo number density would be
\begin{equation}
	\frac{\diff n}{\diff\ln M}=\frac{\bar\rho}{M}\frac{\diff f}{\diff\ln M}.
\end{equation}
Here, and for the bulk of this work, we use the $M_{200}$ halo mass definition, which is the mass of the sphere (centered on the minimum of the gravitational potential) which encloses average density $200\bar\rho$.

\subsection{Prompt cusps and initial peaks}\label{sec:peakcusp}

For the main analysis in this work, we take advantage of the tight connection between prompt cusps and the initial density field.
Each prompt cusp arises from the collapse of an initial density peak. According to \citet{2019PhRvD.100b3523D}, \citet{2023MNRAS.518.3509D}, and \citet{2024MNRAS.52710802O}, the properties of a cusp are related to the collapse time $\tcoll$ of the peak (or scale factor $\acoll$) and the characteristic length scale
\begin{align}\label{L}
	L=|\delta/\nabla^2\delta|^{1/2}
\end{align}
of the peak, where $\delta=(\rho-\bar\rho)/\bar\rho$ is the local maximum (linear-theory) density contrast and $\nabla^2\delta$ is the Laplacian of the density contrast at the site of the local maximum. The prompt cusp that forms from this peak has density profile $\rho=A r^{-1.5}$ within a total mass $m$, where
\begin{align}\label{cusp}
	A &=\alpha \left[\bar\rho L^{1.5}\right]_{\rm coll},
	\nonumber\\
	m &=\beta \bar\rho L^3.
\end{align}
Here $\alpha\simeq24$ and $\beta\simeq7.3$ are numerical coefficients; $\alpha$ is matched to simulation results while $\beta$ comes from a theoretical argument \citep[see][]{2023MNRAS.518.3509D}. Also, the subscript ``coll'' here means that $\bar\rho$ and $L$ are evaluated at the time that the peak collapses, which we evaluate using the \citet{2001MNRAS.323....1S} ellipsoidal collapse threshold (which depends on the local tidal shear).

\subsection{Tracking prompt cusps}\label{sec:tracking}

\begin{figure}
	\centering
	\includegraphics[width=\columnwidth]{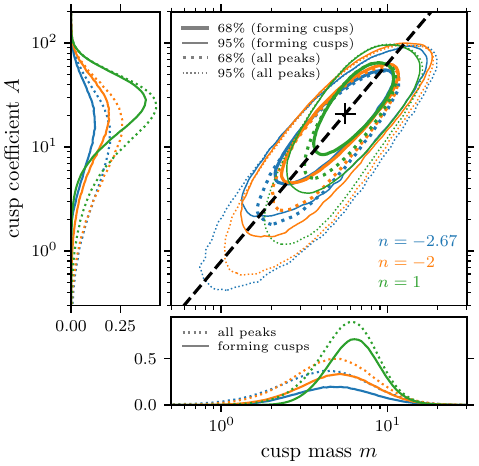}
	\caption{Distribution of density peaks in the initial conditions of the three simulations (different colors) in terms of the resulting cusp mass $m$ and cusp coefficient $A$. The dotted curves show the overall distribution of peaks, while the solid curves show the subset of peaks that are interpreted as forming cusps in the simulations, such that we are able to assign them as halo centers. The main panel shows the joint distribution of $m$ and $A$ (as contours enclosing 68 percent and 95 percent of the distribution), while the left and bottom panels show the marginalized distributions of $m$ and $A$ separately (with measure $\ln A$ or $\ln m$). The diagonal dashed line indicates $A=0.8m^{1.9}$ (equation~\ref{A-m}), which matches the solid-line distributions reasonably well (see also figure~\ref{fig:cusps_A-m}). The cross marks the characteristic properties $(\mchar,\Achar)|_{\sigma_0=1}$ of cusps forming at the time when $\sigma_0=1$, according to equations~(\ref{mAchar}); it conveniently lies near the center of the cusp distribution. See table~\ref{tab:units} for units.}
	\label{fig:peaks_A-m}
\end{figure}

In the initial density field of each simulation, we identify every local maximum and use Fourier methods to evaluate $\nabla^2\delta$ and the tidal shear \citep[e.g.][]{2019PhRvD.100b3523D}. We evaluate the length scale $L=|\delta/\nabla^2\delta|^{1/2}$, and we use the tidal shear to evaluate the collapse time using the fitting function of \citet{2001MNRAS.323....1S}. With these results, we can use equations~(\ref{cusp}) to associate each peak with a prompt cusp with properties $(m,A)$. The dotted curves in figure~\ref{fig:peaks_A-m} show the distribution of associated cusp parameters $(m,A)$ for all initial density peaks.

We wish to track the material associated with these peaks over time in order to determine where the resulting cusp lies. For this purpose, we identify the 7 nearest particles to each density peak in the initial conditions, and we track the halo membership of these particles over time.
If at least 4 of the 7 particles belong to the same halo, and the time is later than the predicted collapse time of the initial peak, then the resulting cusp is deemed to belong to that halo. We checked that changing this threshold to 5 or 6 of the 7 particles has a negligible impact on the results.

However, we are interested in central cusps. Each halo can have only one central cusp; all other cusps are subhalo cusps to the extent that they survive \citep[see][]{2023JCAP...10..008D}.
Of all of the cusps that belong to a halo according to our procedure, the central cusp is preferred to be one that was already the central cusp of a halo in the previous time step. 
This consideration reflects that we do not expect a subhalo's central cusp to replace the main halo's central cusp.
If multiple cusps were halo centers in the previous time step, then the central cusp of the most massive halo is favored.
This preference is intended to capture the effect of halo mergers; the largest halo in a merger event is taken to be the new main halo.\footnote{The more massive halo's tidal forces should disrupt the structure of the less massive halo before the reverse happens. However, since density factors into tidal susceptibility, another possibility might be to favor the denser cusp in such a merger. We tried this option as well, and it did not significantly change the results. As we will see, more massive halos tend to have denser cusps anyway.} The time step for this procedure is $\Delta\ln a\simeq 0.034$.

In the event that no cusp was a halo center in the previous time step, we assume that we are forming a new halo with a new central cusp. If there are multiple candidate cusps, then the central cusp is taken to be the one with the earliest predicted collapse time.\footnote{This approach of tracking halo membership of peak-associated particles is similar to the cusp survival analyses of \citet[][Appendix~B]{2023JCAP...10..008D} and \citet[][Appendix~C]{2024arXiv240318893G}. However, unlike those works, here we make the analysis more transparent and robust by only considering FOF groups (field halos) and not \textsc{subfind-hbt} objects (subhalos).}

The solid curves in figure~\ref{fig:peaks_A-m} show the distribution of cusp properties for peaks that were ever associated with halo-center cusps through this procedure. Note that prompt cusp formation only occurs in the field (and not in halos), so every cusp must have been the center of a field halo (as opposed to a subhalo) at some point. Therefore, these distributions can be interpreted as the overall distribution of prompt cusps.\footnote{However, FOF groups are only identified with a minimum of 32 particles, so this procedure may miss some cusps whose halos merged onto larger systems before they could grow above this resolution limit.}

\subsection{Census of bound objects}

An interesting feature of our cusp-tracking procedure is that it allows us to easily distinguish halos from discreteness artifacts. 
$N$-body simulations are known to produce an abundance of spurious halos on scales comparable to or smaller than the cutoff scale in the matter power spectrum
\citep[e.g.][]{2007MNRAS.380...93W,2013MNRAS.434.3337A,2014MNRAS.439..300L,2020MNRAS.495.4943S,2022MNRAS.509.1703S,2024MNRAS.52710802O}.
These objects typically arise through fragmentation of filamentary structures.
The low-mass ends of the FOF group mass functions shown in figure~\ref{fig:massfunction} (dashed curves) include contributions from these artificial fragments.

Since halo formation begins at local maxima in the initial density \citep{2019PhRvD.100b3523D,2023MNRAS.518.3509D}, we expect that halos with central cusps assigned through our analysis are real halos. Meanwhile, halos without assigned cusps are expected to be spurious, since they did not originate from density maxima.
Based on these considerations, the solid curves in figure~\ref{fig:massfunction} show the mass functions for halos that are expected to be real.
Consistently with previous works \citep[e.g.][]{2013MNRAS.434.3337A,2014MNRAS.439..300L}, we find that spurious fragments considerably inflated the abundance of halos on the lowest mass scales.

\begin{figure}
	\centering
	\includegraphics[width=\columnwidth]{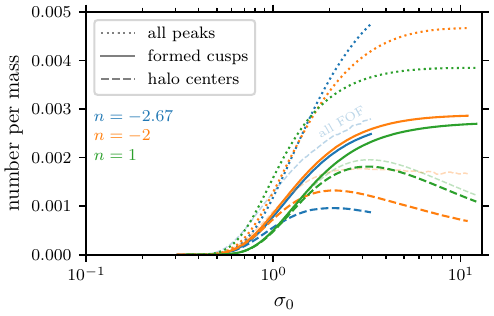}
	\caption{Total number of objects per mass of dark matter as a function of time (parameterized by the rms density contrast $\sigma_0$) within the different simulations (different colors). The dotted curves show the total number of initial density peaks that would have been predicted to collapse. The solid curves show the number that have ever been uniquely associated with field halos in the simulation. This number can be regarded as the cumulative number of cusps that ever form, and it is interestingly universal across cosmologies. The dashed curves show how many cusps can be instantaneously identified as the centers of field halos; this number decreases over time due to halo mergers. For comparison, the faint dashed curves show the total number of FOF groups identified in the simulation, which is larger because of artificial fragmentation. See table~\ref{tab:units} for units.}
	\label{fig:count}
\end{figure}

We can use these results to discuss the number of bound objects in the universe.
Figure~\ref{fig:count} shows the total number of cusps over time.
Per mass of dark matter, the number of peaks in the initial conditions varies considerably between the different cosmologies, with $n=-2.67$ having the highest density of peaks and $n=1$ the lowest.
However, the number of peaks that actually collapse to form cusps according to our analysis (solid curves) turns out to be nearly independent of cosmology in our natural units. At late times $\sigma_0\gg 1$, it converges universally to about one object per mass
\begin{align}\label{count}
	M_*= 350 \bar\rho(\sigma_0/\sigma_2)^{3/2}
\end{align}
of dark matter. Here we appropriately include the mass unit from table~\ref{tab:units}. The number $1/M_*$ can be regarded as an upper bound on the total number of prompt cusps per mass of dark matter. Equivalently, it is an upper bound on the total number of halos and subhalos, since all such objects originate as prompt cusps.

The focus of this work, however, is on the central cusps of field halos (as opposed to subhalos), and the abundance of field halos is suppressed over time by halo mergers.
The dashed curves in figure~\ref{fig:count} show that, at any given time and in the same natural units, the number of field halos (identified with central cusps through our procedure) is lowest for $n=-2.67$ and highest for $n=1$.
For comparison, we also show the total number of FOF groups identified in the simulations (faint dashed curves). Due to artificial fragmentation, as discussed above, this number can moderately exceed the number of true halos.

\section{Properties of prompt cusps}\label{sec:cusps}

\begin{figure*}
	\centering
	\includegraphics[width=\linewidth]{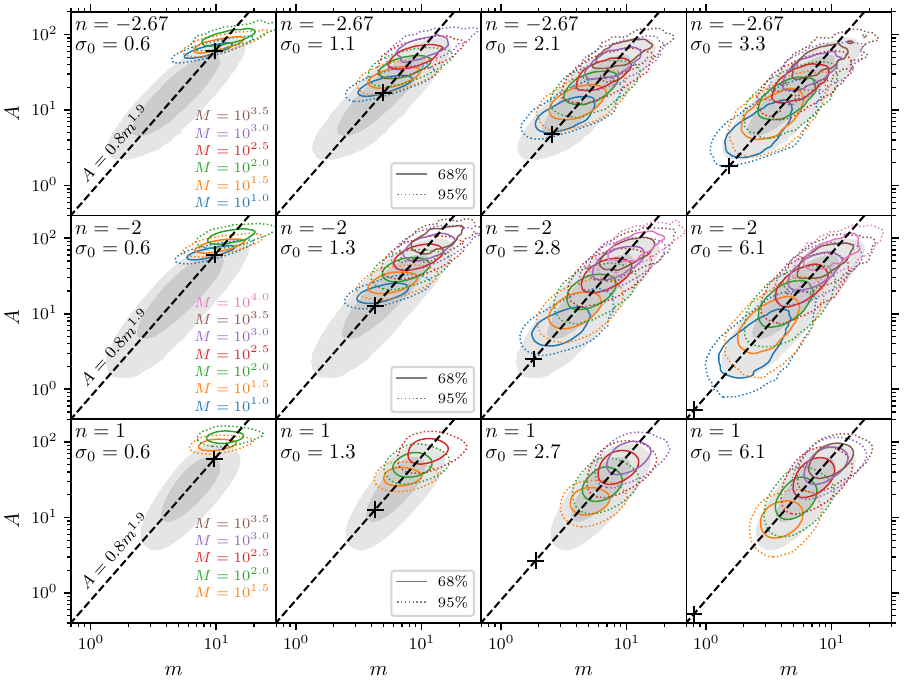}
	\caption{Distribution of halo-center cusps in terms of the cusp mass $m$ and cusp coefficient $A$. We separate the distribution by simulation (different rows), time (different columns, parameterized by $\sigma_0$), and halo mass $M$ (different colors). Solid curves enclose 68 percent of the distribution, while dotted curves enclose 95 percent. Cusps are included if their halo mass $M$ lies within a factor of $10^{1/4}\simeq 1.8$ of the labeled mass, and we only plot cusp distributions in halo mass bins containing at least 500 halos. The diagonal dashed line indicates equation~(\ref{A-m}), which crosses the centers of all of the cusp distributions reasonably accurately. The crosses indicate the characteristic properties of newly forming cusps at each time according to equations~(\ref{mAchar}). The faint gray shading indicates the overall distribution of cusps (same as the solid curves in figure~\ref{fig:peaks_A-m}). See table~\ref{tab:units} for units.}
	\label{fig:cusps_A-m}
\end{figure*}

We now quantify the properties of the prompt cusps that are identified as halo centers in our analysis.
Figure~\ref{fig:peaks_A-m} shows that in terms of cusp coefficient $A$ and mass $m$, the distribution of cusps that are ever identified as halo centers (solid contours) approximately follows
\begin{align}\label{A-m}
	\frac{A}{\bar\rho\sigma_0^{9/4}\sigma_2^{-3/4}} &\simeq C \left[\frac{m}{\bar\rho(\sigma_0/\sigma_2)^{3/2}}\right]^p,
\end{align}
with $C\simeq 0.8$ and $p\simeq 1.9$ (dashed line), independently of the simulated cosmology.
Note the inclusion of appropriate units from table~\ref{tab:units}.
More importantly, figure~\ref{fig:cusps_A-m} shows that equation~(\ref{A-m}) separately holds for halo-center cusps selected at specific times and for specific host halo masses. Although there is scatter in the relationship, that scatter is not significantly correlated with any parameters of interest for this work. Overall, this $A$-$m$ relation appears to be reasonably universal.

A priori, there is little reason to anticipate the existence of a cosmology-independent relationship of the form in equation~(\ref{A-m}). Such a relationship can be partially explained by the statistics of peaks in Gaussian random fields; it would be associated with a statistical relationship between a peak's normalized height $\delta/\sigma_0$ and its normalized curvature $\nabla^2\delta/\sigma_2$. However, the statistics of peaks can only explain the overall distribution of initial peaks (dotted curves in figure~\ref{fig:peaks_A-m}). They cannot explain the overall distribution of cusps (solid curves in figure~\ref{fig:peaks_A-m}), let alone the distributions of halo-center cusps at different times and halo masses (figure~\ref{fig:cusps_A-m}). Thus, the physics of hierarchical gravitational clustering and the structure of the initial conditions beyond the density peaks also play a role in the $A$-$m$ relationship. Additionally, the statistics of peaks do not predict an $A$-$m$ relationship that is independent of the power spectrum (or even of power-law form). Due to all of these complications, we expect that equation~(\ref{A-m}) is not truly independent of the power spectrum and may vary slightly with different cosmologies. Nevertheless, since the variation between cosmologies appears to be small, we will approximate that this equation holds in the same form for all cosmologies.

\subsection{Characteristic properties of young cusps}

The value of the $A$-$m$ relationship in equation~(\ref{A-m}) is that, in conjunction with the peak-cusp connection in equations~(\ref{cusp}), it makes a closed system of equations, which can be solved for $A$ and $m$ to yield
\begin{align}\label{mAchar}
	\Achar&=
	\alpha \left(\frac{\alpha}{C\beta^p}\right)^\frac{1}{2p-1}
	\left[
	\bar\rho\sigma_0^{\frac{9-6p}{4-8p}}\sigma_2^{-3/4}
	\right]_\mathrm{coll},
	\nonumber\\
	\mchar&=
	\beta \left(\frac{\alpha}{C\beta^p}\right)^\frac{2}{2p-1}
	\left[
	\bar\rho\sigma_0^{\frac{9-6p}{2-4p}}\sigma_2^{-3/2}
	\right]_\mathrm{coll}
\end{align}
as functions of the collapse time only (via $\bar\rho$, $\sigma_0$, and $\sigma_2$).
Numerically, the prefactors are $20.9$ for $\Achar$ and $5.57$ for $\mchar$.
At each time, these $\Achar$ and $\mchar$ may be interpreted as the characteristic properties of newly formed, ``young'' cusps.
The crosses in figure~\ref{fig:cusps_A-m} show the value of $(\mchar,\Achar)$ at each time according to equations~(\ref{mAchar}).
As we might expect, this ``characteristic young cusp'' is usually close in $m$ and $A$ to the central cusps of the smallest halos. An exception is when $(\mchar,\Achar)$ lies far outside the overall cusp distribution, as in the lower right panels; the physical interpretation here is that the epoch is late enough that cusps are no longer forming in significant numbers.
We also point out that generally all halos have masses $M\gtrsim\mchar$, as we should expect since halos form around cusps.

The cross in figure~\ref{fig:peaks_A-m} marks $(\mchar,\Achar)|_{\sigma_0=1}$, the characteristic cusp properties for the collapse time corresponding to $\sigma_0=1$. This point turns out to lie near the center of the overall cusp distribution. This outcome is coincidental; it simply means that the typical relative peak height $\delta/\sigma_0$ (for peaks that form cusps) is comparable to the typical ellipsoidal collapse threshold. Nevertheless, these $(\mchar,\Achar)|_{\sigma_0=1}$ are useful reference values of $A$ and $m$, since they depend only on the cosmological scenario.

\subsection{Halo mergers and cusp evolution}\label{sec:mergers}

So far, we have only discussed the properties $(m,A)$ with which the prompt cusps form.
Cusps generally remain static over time even as halos grow around them \citep{2023MNRAS.518.3509D}. However, previous works have noted that major merger events -- that is, merging of halos of comparable mass -- can induce evolution of the central $\rho\propto r^{-1.5}$ density cusps \citep{2016MNRAS.461.3385O,2017MNRAS.471.4687A}.
The physical mechanism responsible for this evolution is that dynamical friction efficiently drains orbital energy from high-mass subhalos, causing them to rapidly sink to the center of a host halo. In a major merger, material from the smaller halo can quickly reach the center of the larger halo, adding new mass and energy that could either raise or lower the density of the central cusp, depending on the configuration \citep{2016MNRAS.461.3385O,2017MNRAS.471.4687A}.

\begin{figure}
	\centering
	\includegraphics[width=\columnwidth]{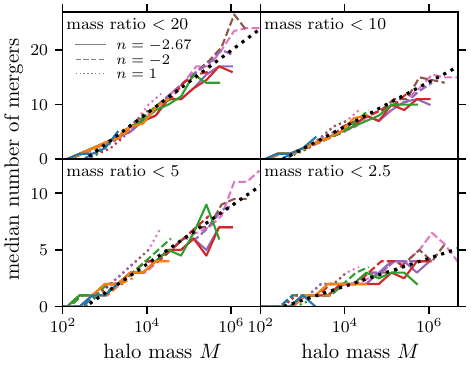}
	\caption{Median number of mergers experienced by a halo of mass $M$ with mass ratio below 20, 10, 5, or 2.5 (different panels). Different line styles represent the three simulations, while colors correspond to different times with the same meaning as in figure~\ref{fig:massfunction}. We use halo mass bins of width $\Delta\ln M=0.7$; units are in table~\ref{tab:units}. Across all cosmologies and times, the number of mergers is well approximated by equation~(\ref{Nmerge}), shown here with the thick black dotted line.}
	\label{fig:mergers}
\end{figure}

We now explore how many major mergers each halo experiences.
This number is extracted straightforwardly from our cusp-tracking procedure in section~\ref{sec:tracking}.
For two different mass-ratio thresholds, figure~\ref{fig:mergers} shows the median number of major mergers as a function of halo mass.
The result is surprisingly independent both of time and of cosmology. Indeed, for a halo of mass $M$, the median number $N_{\mathrm{ratio}<x}(M)$ of past mergers with mass ratios lower than $x$ is well approximated by
\begin{equation}\label{Nmerge}
	N_{\mathrm{ratio}<x}(M)\simeq
	0.6\left(\ln x\right)^{1.3}
	\ln(M/M_*)
\end{equation}
if $M>M_*$ and $N_{\mathrm{ratio}<x}(M)\simeq 0$
otherwise, where $M_*= 350 \bar\rho(\sigma_0/\sigma_2)^{3/2}$ as in equation~(\ref{count}).
Here we define $x$ as the ratio of the larger to the smaller mass, so lower mass ratios (closer to 1) correspond to ``more major'' mergers. Figure~\ref{fig:mergers} shows that this expression (black dotted line) is accurate for the range $1 < x \lesssim 20$.

Note that the appearance of the same constant $M_*= 330 \bar\rho(\sigma_0/\sigma_2)^{3/2}$ in equations (\ref{count}) and~(\ref{Nmerge}) is appropriate.
Equation~(\ref{Nmerge}) means that a halo is not expected to undergo a merger until it has grown to a mass of about $M_*$, while according to equation~(\ref{count}), $M_*$ is precisely the amount of cosmological mass inside which one prompt cusp is expected to form.
Consequently, these equations imply that a halo does not accrete other halos (on average) until it has grown to a mass in which more than one halo would be expected to form.

Figure~\ref{fig:mergers} shows that sufficiently massive halos can have undergone $\mathcal{O}(10)$ major mergers. However, for cusp evolution, the simulations of \citet{2023MNRAS.518.3509D} suggest that it is also important to consider the halo mass when each merger took place.
If the central prompt cusp comprises a large fraction of the halo, then infalling subhalos can reach the cusp before they are significantly disrupted by the host's tidal forces. For example, this is the case in the idealized simulations of \citet{2016MNRAS.461.3385O} and \citet{2017MNRAS.471.4687A}, which considered ``bare'' prompt cusps without halo envelopes.\footnote{\citet{2016MNRAS.461.3385O} used a density profile that is equivalent to equation~(\ref{cuspNFW0}) with $y=1$, while \citet{2017MNRAS.471.4687A} adopted $\rho\propto r^{-1.5}$ power-law density profiles with a sharper outer truncation.}
On the other hand, if the central cusp is embedded deep inside a broader halo, then the tidal forces induced by the halo would greatly suppress the mass of any infalling subhalo before it could reach the central cusp, and so the effect on the cusp would be lessened.
In cosmological simulations, \citet{2023MNRAS.518.3509D} found two instances where mergers with mass ratios $x\sim 4$ resulted in an approximately $10$ percent reduction in the density of the central cusp. In these cases, the halo mass $M$ at the time of the merger was about 50 times the mass $m$ of the initial prompt cusp.
Meanwhile, \citet{2023MNRAS.518.3509D} also found an instance where a merger with mass ratio $x\sim 3$ had no effect on the central cusp. In this case, the halo mass at the time of the merger was about $M\sim 200m$. These outcomes suggest that $M\lesssim 10^2 m$ may be required for major mergers to alter the central cusp, although a more systematic study of merger outcomes is needed.

Figure~\ref{fig:mergers} shows that major mergers are rare for halos with $M\lesssim 10^2\bar\rho(\sigma_0/\sigma_2)^{3/2}$. Since prompt cusps generally have $m\lesssim 10\bar\rho(\sigma_0/\sigma_2)^{3/2}$ (see figure~\ref{fig:peaks_A-m}), major mergers between ``bare'' cusps \citep[as simulated by][]{2016MNRAS.461.3385O,2017MNRAS.471.4687A} are highly atypical.
Instead, halos typically experience about one major merger by the time they grow to the much larger mass of $M\simeq 10^3\bar\rho(\sigma_0/\sigma_2)^{3/2}$ (depending on the threshold mass ratio).
But $10^3\bar\rho(\sigma_0/\sigma_2)^{3/2}\sim 10^2 m$ for typical cusp masses $m$, which is about the threshold above which mergers are not expected to affect the central cusp, per the discussion above. Consequently, we can expect that halos each experience about 1 merger that affects the central cusp at around the $10$ percent level.\footnote{However, the results of \cite{2024MNRAS.52710802O} suggest that mergers are more important for scenarios with shallow spectral cutoffs. Shallow cutoffs can arise from free streaming if the dark matter velocity distribution is bimodal, such as in multicomponent scenarios \citep[e.g.][]{2017JCAP...12..013B,2020PhRvD.101l3511D,2022PhRvD.106h3506D}.}
Due to the limited degree to which cusp evolution in halo mergers has been studied, we do not attempt to directly incorporate this effect into our modeling.
Instead, we conclude for now that halo mergers give rise to a roughly $10$ percent uncertainty in the cusp coefficients $A$.

\section{The cusp-halo relation}\label{sec:cusphalo}

\begin{figure*}
	\centering
	\includegraphics[width=\linewidth]{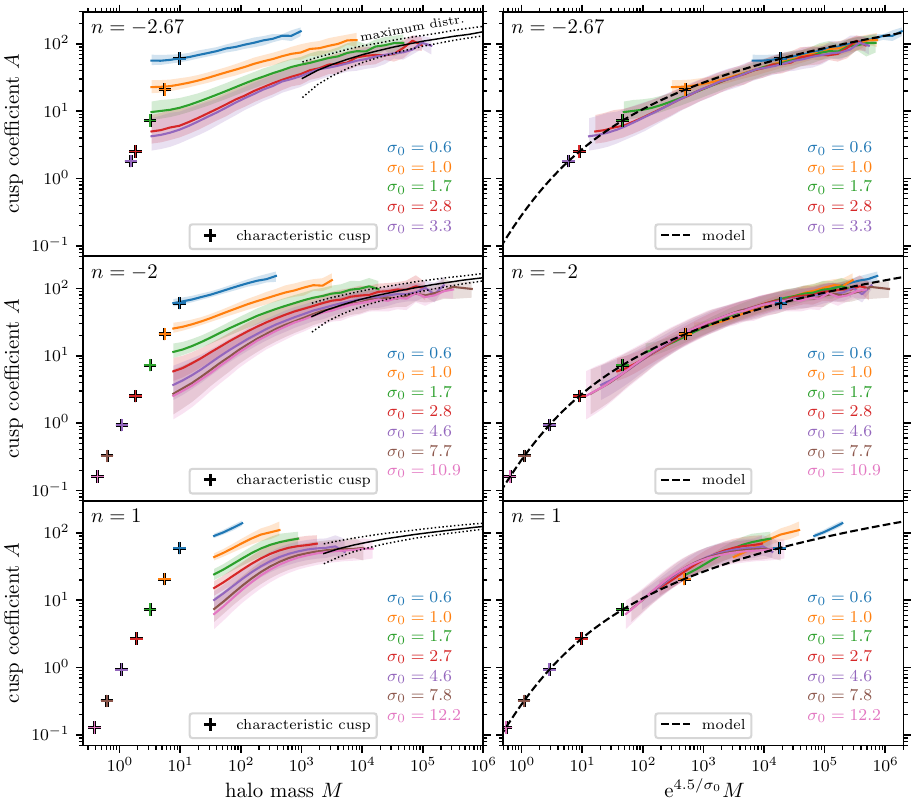}
	\caption{Coefficients $A$ of the central cusps of halos of mass $M$. Colored curves in the left-hand panels show the median $A$ in each halo mass bin, and we shade between the 16th and 84th percentiles. Halo mass bins have width $\Delta\ln M\simeq 0.35$, and we only consider masses $M>32m_\mathrm{p}$, where $m_\mathrm{p}$ is the simulation particle mass. Different panels correspond to simulations of different cosmologies, while different colors represent different times (parameterized by $\sigma_0$). Units are in table~\ref{tab:units}. For comparison, the black curves in the left-hand panels mark the median and 16th/84th percentiles of the distribution of the maximum cusp coefficient $A$ for initial density peaks in an arbitrary cosmological volume of mass $M$.
	The right-hand panels show that with halo masses scaled by $1/\chi(\sigma_0)=\e^{4.5/\sigma_0}$, all of the $A$-$M$ curves overlap. This outcome enables the construction of a cusp-halo relation (dashed line) using the characteristic properties of young cusps in each epoch, which we show as crosses in all panels.}
	\label{fig:A-M}
\end{figure*}

\begin{figure*}
\centering
\includegraphics[width=\linewidth]{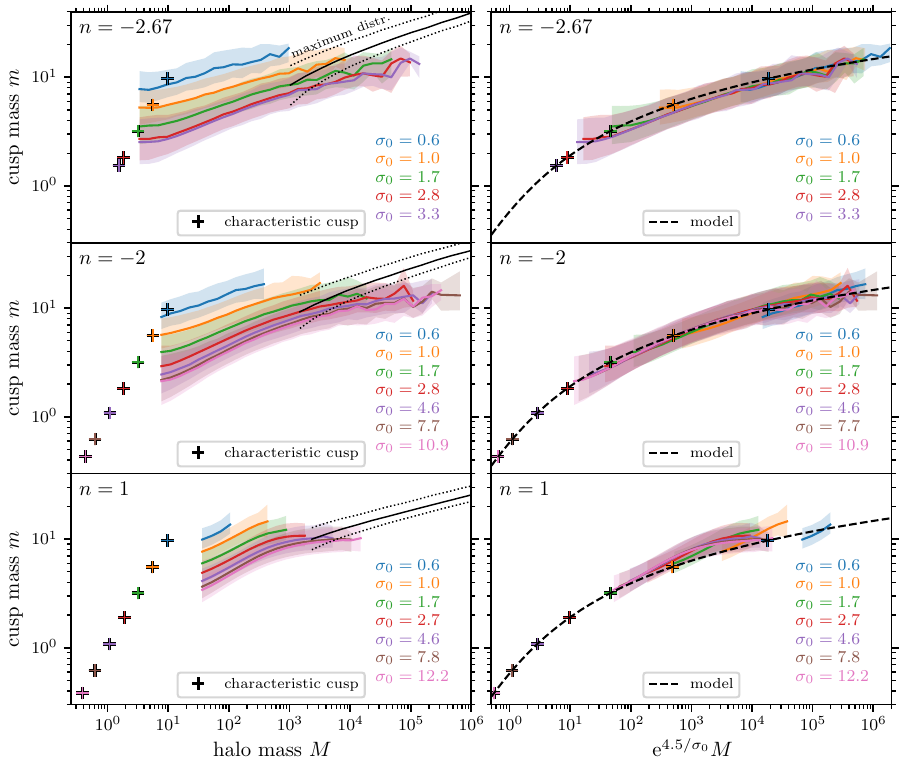}
\caption{Similar to figure~\ref{fig:A-M} but showing the cusp mass $m$, instead of the coefficient $A$, as a function of halo mass $M$. Note that the characteristic ``young'' cusps (crosses) lie on the $m=M$ line by construction. See table~\ref{tab:units} for units.}
\label{fig:m-M}
\end{figure*}

We are now prepared to develop the cusp-halo relation. 
The left-hand panels of figures \ref{fig:A-M} and~\ref{fig:m-M} show how the central cusp $A$ and $m$, respectively, depend on the halo mass $M$.
At any fixed time, more massive halos tend to have denser and more massive central cusps. Meanwhile, in any fixed halo mass bin, the central cusps become less dense and less massive over time.

It is interesting to note that, at late times $\sigma_0\gg 1$ and large halo masses $M$, the distribution of cusp coefficients $A$ lies close to the distribution of the maximum predicted $A$ for density peaks in a random cosmological volume of mass $M$. We evaluate this maximum distribution by randomly drawing a fraction $M/M_\mathrm{box}$ of all of the initial peaks in the simulation volume, where $M_\mathrm{box}$ is the total mass in the simulation, and recording the largest $A$ in this sample. By repeating this process $10^4$ times for each mass $M$, we arrive at the distributions shown in black in the left-hand panels of figure~\ref{fig:A-M}.
Physically, this outcome seems to reflect that the cusps of highest $A$ typically form earliest (see equation~\ref{mAchar}) and thus have time to accrete the largest halos. In halo mergers, the higher-$A$ cusp thus typically occupies the larger halo and becomes the central cusp of the descendant halo.
Note however that the same outcome does not apply to cusp masses $m$; the central cusp of a halo of mass $M$ tends to be significantly lower than the largest cusp mass $m$ for initial peaks in a cosmological volume of mass $M$ (black curves in the left-hand panels of figure~\ref{fig:m-M}).

\subsection{Functional form of the cusp-halo relation}\label{sec:cusphaloform}

Physically, one should picture the $A$-$M$ and $m$-$M$ curves shifting to the right over time; halos become more massive over time while their central cusps remain static.
Although halo growth can in general be a complicated function of mass and time, remarkably we find that a simple, universal scaling factor
\begin{align}\label{massgrowth}
	\chi(\sigma_0)&=\e^{-\kappa/\sigma_0}
\end{align}
works well with $\kappa\simeq 4.5$, such that the cusp properties $m$ and $A$ are nearly universal as a function of $M/\chi(\sigma_0)$. The right-hand panels of figures \ref{fig:A-M} and~\ref{fig:m-M} demonstrate this feature.
What this outcome means is that for fixed central cusp properties ($m$,$A$), the masses of halos with those cusps increase over time precisely in proportion with $\chi(\sigma_0)$ (recalling that $\sigma_0$ increases over time). Since the number of central cusps decreases over time as halos merge, this outcome does not necessarily require that halos themselves grow in mass proportionally with equation~(\ref{massgrowth}).
However, we show in appendix~\ref{sec:concentration} that equation~(\ref{massgrowth}) is indeed a reasonably accurate, universal description of the median mass accretion histories of sufficiently low-mass halos.

Now consider the characteristic ``young'' cusp properties ($\mchar$,$\Achar$) derived in section~\ref{sec:cusps}. We include these characteristic cusps in figures \ref{fig:A-M} and~\ref{fig:m-M}, interpreting them as belonging to halos of the same mass $M=\mchar$.
In the cases where $\mchar$ is above the halo mass resolution limit, the characteristic cusp lies on or close to the cusp-halo ($A$-$M$ and $m$-$M$) relations.
More generally, when scaled in halo mass by equation~(\ref{massgrowth}) in the right-hand panels, these characteristic cusps lie on (or close to) the same universal cusp-halo relation.

The very natural interpretation of this outcome is that halos grow from an early phase in which they are cusp alone. Then the halo mass $M$ should be related to the cusp mass according to the mass scaling factor in equation~(\ref{massgrowth}), leading to
\begin{align}\label{Mgrowth}
	M &= \chi\left[\mchar/\chi\right]_\mathrm{coll},
\end{align}
where the quantity in brackets is evaluated at the initial collapse time of the halo's cusp. Substituting the expressions for $\chi$ (equation~\ref{massgrowth}) and the characteristic cusp mass $\mchar$ (equation~\ref{mAchar}) yields
\begin{align}\label{scoll_eq}
	\left[\sigma_0^{\frac{3}{2p-1}}\chi(\sigma_0)\right]_\mathrm{coll}\!\!
	&=\left(\frac{\alpha^2}{\beta C^2}\!\right)^{\!\!\frac{1}{2p-1}\!}\frac{\bar\rho(\sigma_0/\sigma_2)^{3/2}}{M}\chi(\sigma_0),
\end{align}
which can be solved explicitly for the collapse time if $\chi(\sigma_0)=\e^{-\kappa/\sigma_0}$, yielding
\begin{align}\label{scoll}
	\sigma_0|_\mathrm{coll} &=
	\frac{(2p-1)\kappa/3}{
		\lambertW\!\!\left\{\!
		\frac{(2p-1)\kappa}{3}\!
		\left(\frac{\beta C^2}{\alpha^2}\right)^{\!1/3\!}
		\left[\frac{\e^{\kappa/\sigma_0}M}{\bar\rho(\sigma_0/\sigma_2)^{3/2}}\right]^{\frac{2p-1}{3}}\!\right\}\!
	}.
\end{align}
Here $\lambertW$ is the Lambert W function (also known as the product logarithm), defined by $\lambertW(x)\e^{\lambertW(x)}=x$. Table~\ref{tab:parameters} reviews the value and meaning of each of the model parameters that appear in these expressions.

\begin{table}
	\centering
	\caption{Parameters that enter into the cusp-halo relation.}
	\label{tab:parameters}
	\begin{tabular}{ccc}
		\hline
		param. & description & equation \\
		\hline
		$\alpha\simeq 24$ & coef. of $A$ in peak-cusp connection & (\ref{cusp}) \\
		$\beta\simeq 7.3$ & coef. of $m$ in peak-cusp connection & (\ref{cusp}) \\
		$C\simeq 0.8$ & coefficient in cusp $A$-$m$ relationship & (\ref{A-m}) \\
		$p\simeq 1.9$ & index in cusp $A$-$m$ relationship & (\ref{A-m}) \\
		$\kappa\simeq 4.5$ & parameter in mass growth factor & (\ref{massgrowth}) \\
		\hline
	\end{tabular}
\end{table}

For a halo of mass $M$, equation~(\ref{scoll}) yields the value of $\sigma_0$ at the initial formation time of the halo's central cusp, which in turn determines the formation time.
Using the $\sigma_j$ (equation~\ref{sigmaj}) and density $\bar\rho$ evaluated at this formation time, equations~(\ref{mAchar}), which give the properties $(m,A)$ of young cusps, may then be evaluated to produce a prediction for the properties of the halo's central cusp. The dashed curves in the right-hand panels of figures \ref{fig:A-M} and~\ref{fig:m-M} show that prediction.
Overall, the cusp-halo relation defined by these equations works reasonably well across all of the cosmologies, times, and halo masses that we have considered. We emphasize that given how broadly our simulations represent the possible matter power spectra, this relation can be expected to hold generally at a comparable degree of precision.

There are nevertheless some systematic discrepancies. The cusp-halo relation defined by equations (\ref{mAchar}) and~(\ref{scoll}) slightly overestimates the cusp $A$ and $m$ in the $n=-2.67$ simulation and underestimates them in the $n=1$ simulation.
The cusp-halo relation is built on two main ingredients that we have assumed to be independent of cosmology: the cusp $A$-$m$ relationship in equation~(\ref{A-m}) and the mass scaling factor $\chi(\sigma_0)$ in equation~(\ref{massgrowth}). It is likely that neither of these is truly independent of cosmology, but we leave for future work the possibility of tuning the cusp-halo relation for specific cosmologies. The universal form that we have presented should be understood to be accurate to within around 30 percent in predictions of the median coefficient $A$ or mass $m$ of central cusps. For warm dark matter models, which most closely resemble the $n=-2$ simulation, the middle panels of figures \ref{fig:A-M} and~\ref{fig:m-M} suggest that any systematic offset is instead only around 10 percent.

Also, we have only considered simulations of matter-dominated cosmologies. Although prompt cusps are not still forming in significant numbers, the recent onset of dark energy domination in a concordance cosmology could influence the evolution of the mass scaling factor $\chi(\sigma_0)$.
However, in appendix~\ref{sec:concentration}, we find that the mass scaling factor in equation~(\ref{massgrowth}) describes mass accretion histories reasonably well in the warm dark matter simulations of \citet{2016MNRAS.455..318B} (with dark energy), so we expect that the cusp-halo relation that we have presented should remain accurate. In this case we use the $M_{200\mathrm{c}}$ halo mass definition, which is the mass enclosed within the sphere of average density 200 times the critical density.

For a general, potentially cosmology-dependent mass scaling factor $\chi(\sigma_0)$, an explicit mathematical expression for $\sigma_0|_\mathrm{coll}$ might not be possible. However, equation~(\ref{scoll_eq}) remains simple to solve numerically. Mass-dependent $\chi(\sigma_0,M)$ can be handled straightforwardly as well, where $M$ is the final halo mass (at the time when we are evaluating its central cusp).

\subsection{Scatter in cusp properties}\label{sec:cusphaloscatter}

\begin{figure}
	\centering
	\includegraphics[width=\columnwidth]{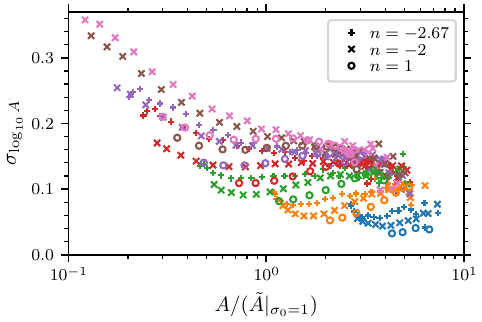}
	\caption{68 percent scatter in the cusp coefficient $A$ at fixed halo mass, time, and cosmology.
	Different markers represent simulations of different cosmologies, while the colors represent time with the same meaning as the colors in figures \ref{fig:A-M} and~\ref{fig:m-M}. We show the scatter as a function of the median $A$, itself expressed in units of the characteristic $A$ of cusps forming at $\sigma_0=1$. For sufficiently large $A\gtrsim \Achar|_{\sigma_0=1}$ and sufficiently late times ($\sigma_0\gtrsim 1$, corresponding to the green, red, purple, brown, or pink markers), the scatter in $A$ is around 0.15 dex.}
	\label{fig:scA}
\end{figure}

\begin{figure}
	\centering
	\includegraphics[width=\columnwidth]{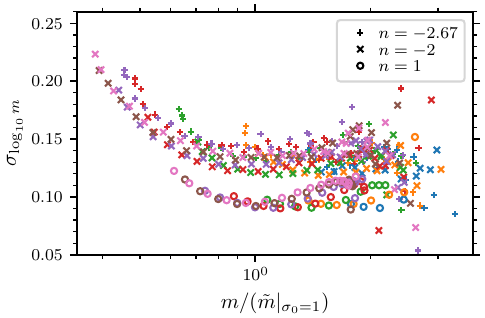}
	\caption{Like figure~\ref{fig:scA} but showing the 68 percent scatter in the cusp mass $m$. For $m\gtrsim \mchar_{\sigma_0=1}$, the scatter in $m$ is around 0.1 to 0.15 dex.}
	\label{fig:scm}
\end{figure}

So far we have focused on the median parameters $A$ and $m$ of central cusps. We now discuss their scatter around the median. For this purpose, we define $\sigma_{\log_{10}A}=\frac{1}{2}\log_{10}(A^\mathrm{(84\%)}/A^\mathrm{(16\%)})$, where $A^\mathrm{(84\%)}$ and $A^\mathrm{(16\%)}$ are the 84th and 16th percentile central cusp coefficients $A$ in each halo mass bin, and $\sigma_{\log_{10}m}$ likewise for central cusp mass $m$. Figure~\ref{fig:scA} shows how the scatter in $A$ depends on $A$, while figure~\ref{fig:scm} shows how the scatter in $m$ depends on $m$.
Generally, as long as $A\gtrsim\tilde A|_{\sigma_0=1}$ and $\sigma_0\gg 1$, the scatter in the cusp coefficient $A$ is around $\sigma_{\log_{10}A}\simeq 0.15$ (0.15 dex, corresponding to a factor of about 1.4). The scatter in $m$ is more variable but is generally around 0.1 to 0.15 dex. Note that figure~\ref{fig:cusps_A-m} shows that cusp $A$ and $m$ are weakly correlated in their scatter, in that (at fixed time and halo mass) there is a slight tendency for cusps of higher $m$ to also have higher $A$, and vice versa.

\begin{figure}
\centering
\includegraphics[width=\columnwidth]{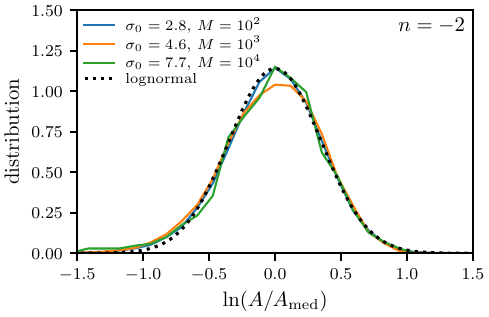}
\caption{Distribution of cusp coefficients $A$ in relation to the median value $A_\mathrm{med}$ at fixed halo mass $M$ and time $\sigma_0$. We consider a few $(M,\sigma)$ (different colors) within the $n=-2$ simulation; halos are selected if their $M_{200}$ lies within a factor of $10^{1/4}\simeq 1.78$ of the nominal $M$. The $A$ are binned in intervals of width $\Delta\ln A \simeq 0.12$. For comparison, the black dotted curve shows a lognormal distribution with a standard deviation of 0.35 e-folds (about 0.15 dex).}
\label{fig:lognormal}
\end{figure}

We show the scatter in more detail in figure~\ref{fig:lognormal}. Within the $n=-2$ simulation, we show the full distributions of $A$ at a few fixed times $\sigma_0$ and halo mass bins. These distributions are closely approximated by the lognormal distribution depicted with a dotted curve. We have separately verified that the scatter in cusp masses $m$ is distributed similarly.
Apparently, the scatter in the cusp-halo distribution is nearly lognormal, and hence the $\sigma_{\log_{10}A}$ and $\sigma_{\log_{10}m}$ discussed above may be interpreted as the standard deviation of the lognormal distribution.

Note that the scatter in the cusp-halo relation arises from a mix of scatter in the properties of young cusps (i.e., deviations from equations~\ref{mAchar}) and scatter in halo assembly histories (i.e., deviations from equation~\ref{Mgrowth}). If a halo's assembly history is already known, then the latter contribution can be suppressed by evaluating the cusp-halo relation early in that history. For example, consider the blue ($\sigma_0=0.6$) and orange ($\sigma_0=1$) points in figure~\ref{fig:scA}, which represent the scatter in cusp coefficients $A$ close to the times that typical halo-center cusps originally form. At these early times, the 68 percent scatter in $A$ is only around 0.06 dex (about 15 percent).

\subsection{Implications for warm dark matter halos}

\begin{figure}
	\centering
	\includegraphics[width=\columnwidth]{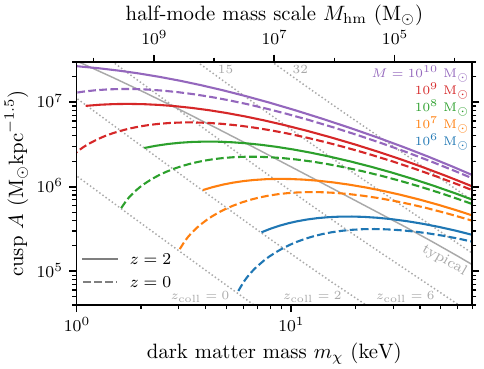}
	\caption{Central cusps of warm dark matter halos at redshift 2 (solid colored curves) or redshift 0 (dashed curves). We show the median cusp coefficient $A$ for field halos of several different masses $M$ (different colors) as a function of the dark matter particle mass $m_\chi$. For comparison, the solid gray diagonal line shows how the typical cusp coefficient (which we define as $\Achar|_{\sigma_0=1}$; see equation~\ref{mAchar}) depends on the dark matter model. For higher $m_\chi$, the free-streaming scale is smaller, so the cusps are smaller on average and have smaller $A$. However, for a halo of set mass, the central cusp's $A$ decreases much more gradually with $m_\chi$ than does the overall distribution throughout the universe. For reference, the dotted diagonals mark the coefficients of cusps forming at fixed redshift; the cusps of more massive halos form earlier. At top we show the half-mode mass scale associated with the warm dark matter model.}
	\label{fig:wdm_cusps}
\end{figure}

We now illustrate some of the implications of this result for warm dark matter models. Figure~\ref{fig:wdm_cusps} shows how the coefficients $A$ of central cusps depend on the dark matter mass $m_\chi$. Here we adopt cosmological parameters from \citet{2020A&A...641A...6P}; we generate a cold dark matter power spectrum using the \textsc{class} code \citep{2011JCAP...07..034B} and apply a free-streaming cutoff according to the prescription of  \citet{2023PhRvD.108d3520V} for a spin 1/2 particle. For reference, we also specify the commonly used ``half-mode'' mass scale $M_\mathrm{hm}$, defined according to the convention of e.g. \citet{2019ApJ...878L..32N}.

For a fixed formation redshift $z_\mathrm{coll}$, the dotted diagonals indicate that cusp coefficients scale as approximately $A\propto m_\chi^{-1.8}$.
Since cusps form earlier for higher $m_\chi$, the solid diagonal shows that typical cusps (defined as those forming when $\sigma_0=1$) have coefficients scaling as approximately $A\propto m_\chi^{-1.3}$.
However, for central cusps of field halos with fixed mass $M$, the colored curves show that cusp coefficients $A$ decrease much more gradually with $m_\chi$.

\citet{2023MNRAS.522L..78D} previously analyzed how observations of dwarf galaxies in the Local Group could constrain warm dark matter if those galaxies contained prompt cusps that formed at $z_\mathrm{coll}>6$ but are otherwise drawn from the overall distribution of density peaks in the initial conditions. But many of the Milky Way satellites should have had halo masses around $10^9$ to $10^{10}~\Msol$ prior to their accretion onto the Milky Way halo. By comparing the red and purple curves in figure~\ref{fig:wdm_cusps} to the solid diagonal and the $z_\mathrm{coll}=6$ dotted diagonal, we can see that for $m_\chi$ above a few keV, the central cusps of these halos should be considerably denser than \citet{2023MNRAS.522L..78D} would have assumed. Hence, \citet{2023MNRAS.522L..78D} considerably underestimated the observational prospects of prompt cusps in these models.

\begin{figure}
	\centering
	\includegraphics[width=\columnwidth]{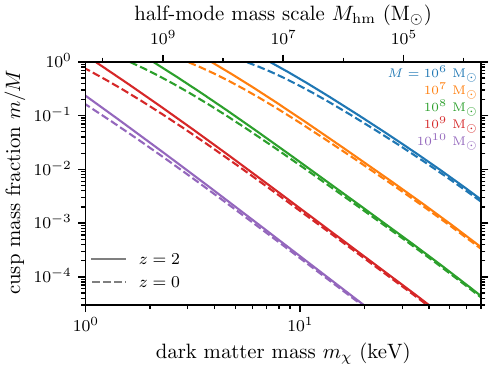}
	\caption{Fraction of halo mass that can be attributed to the central prompt cusp for warm dark matter halos at redshift 2 (solid colored curves) or redshift 0 (dashed curves). As a function of the dark matter particle mass $m_\chi$, we show the ratio $m/M$, where $m$ is the median cusp mass, for field halos of several different masses $M$ (different colors). For higher $m_\chi$, the free-streaming scale is smaller, so the cusps are less massive on average. At top we show the half-mode mass scale associated with the warm dark matter model.}
	\label{fig:wdm_cusps_mass}
\end{figure}

Figure~\ref{fig:wdm_cusps_mass} shows similarly how cusp masses vary with dark matter mass $m_\chi$. Here we show the fraction $m/M$ of the total halo mass that can be attributed to the prompt cusp. For small $m/M\ll 0.1$, the mass fraction scales as approximately $m/M\propto m_\chi^{-3}$, but the scaling can be considerably more gradual when $m/M\gtrsim 0.1$.
We will see in the next section, however, that cusp masses $m$ obtained through the cusp-halo relation are not directly relevant to halo structures. The key role of the cusp mass was in deriving the cusp-halo relation in the first place (per equation~\ref{Mgrowth}). To write down a halo density profile that includes the central prompt cusp, we will see that only the cusp coefficient $A$ is needed.

Finally, we remark that since our simulations only cover about 5 orders of magnitude in halo growth, the cusp-halo relation might not be accurate for $m/M <10^{-5}$. This is not a serious limitation, however, as such a subdominant central cusp would be irrelevant for most purposes.
Also, the cusp-halo relation is directly valid only for field halos (as opposed to subhalos). To determine the central cusp of a subhalo, one should use the mass and redshift at some time before the object accreted onto its host.

\section{A cusp-halo density profile}\label{sec:profile}

\begin{figure*}
\centering
\includegraphics[width=\linewidth]{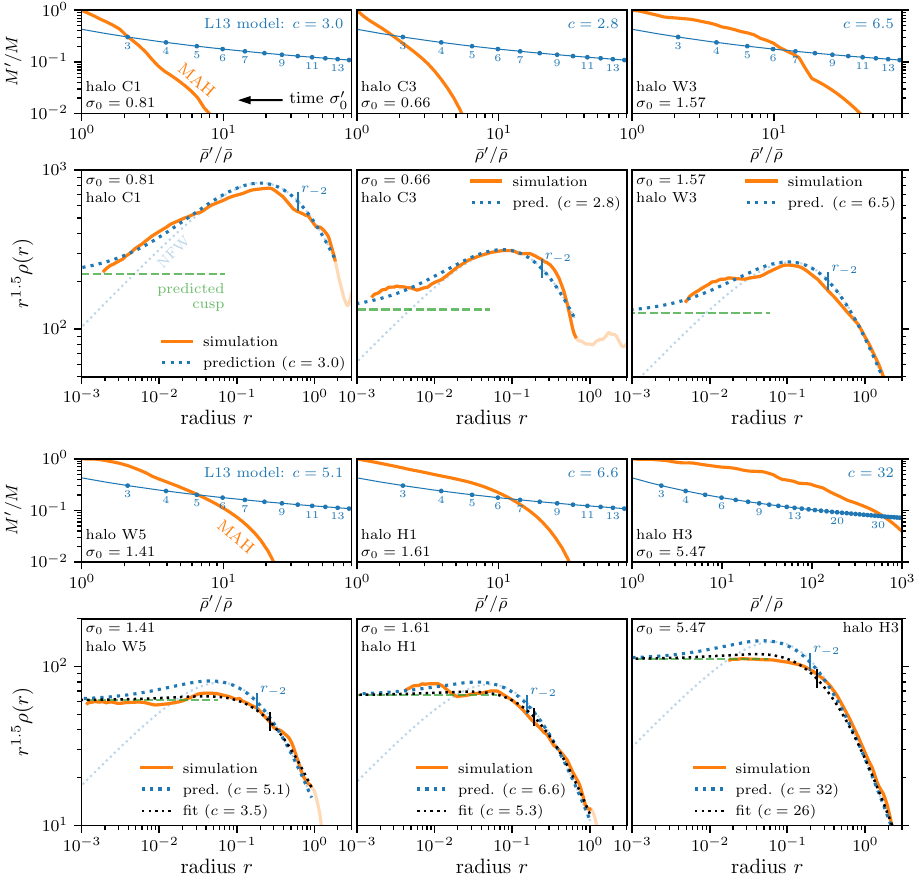}
\caption{Density profiles of halos from the high-resolution subvolume simulations of \cite{2023MNRAS.518.3509D}.
	As a function of radius $r$, the orange curves in the large panels show $r^{1.5}\rho(r)$, where $\rho(r)$ is the spherically averaged density. See table~\ref{tab:units} for units. The horizontal dashed line shows the prompt cusp predicted from the initial conditions, while the dotted curves show ``cusp-NFW'' density profiles (equation~\ref{cuspNFW}); black curves have best-fitting values of the concentration parameter $c$ (lower panels only), while blue curves have $c$ predicted from the mass accretion history. The vertical marks show $r_{-2}=R_{200}/c$. The faint dotted curves show NFW profiles of the same concentration.
	Narrow panels show mass accretion histories of the halos; here we plot $M'/M$, the past mass in units of ``current'' mass (at the time when the density profile is plotted), as a function of $\bar\rho'/\bar\rho$, the past density of the universe in units of the current density (so time advances towards the left). Here the blue lines represent the model of \citet{2013MNRAS.432.1103L} (L13); the point of intersection with the mass accretion history (MAH) yields a prediction for $c$ (blue numbers). We note the rms density contrast $\sigma_0$ (in linear theory) at the time that each density profile is shown.}
\label{fig:profile}
\end{figure*}

We now present a functional form of a density profile that incorporates both a central prompt cusp and the broader halo around it. For this purpose, we analyze the high-resolution subvolumes of the simulations in \citet{2023MNRAS.518.3509D}, which are focused on individual halos.
The large panels of figure~\ref{fig:profile} show (in orange) the density profiles of six of these halos at some particular times.\footnote{The halos and times are chosen to present smooth recent mass accretion histories to facilitate the discussion in section~\ref{sec:cuspNFWacc}.} Halos C1 and C3 come from the $n=-2.67$ simulation, halos W1, W2, and W4 come from the $n=-2$ simulation, and halo H3 comes from the $n=1$ simulation. Density profiles are displayed only for $r>5\epsilon$, where $\epsilon$ is the force softening length, and are shown in faint color for radii $r>R_{200}$.
We also show (with the horizontal dashed line) the prompt cusp that is predicted from the initial conditions in accordance with equations~(\ref{cusp}).

\subsection{The cusp-NFW functional form}\label{sec:cuspNFW}

We propose a ``cusp-NFW'' density profile of the form
\begin{align}\label{cuspNFW}
	\rho(r) &= \frac{\sqrt{y^2+x}}{x^{1.5}(1+x)^2}\rhos, \nonumber\\
	x &= r/\rs, \qquad y = A/(\rhos \rs^{1.5}).
\end{align}
Here $A$ is the cusp coefficient, while $\rhos$ and $\rs$ are characteristic density and radius parameters. In the limit that $A\to 0$, equation~(\ref{cuspNFW}) becomes the classic NFW form, $\rho(r)/\rhos=x^{-1}(1+x)^{-2}$, which transitions from $\rho\propto r^{-1}$ when $r\ll\rs$ to $\rho\propto r^{-3}$ when $r\gg\rs$ and describes halos of arbitrarily cold dark matter accurately \citep{1996ApJ...462..563N,1997ApJ...490..493N}. In the opposite limit that $A\to\rhos \rs^{1.5}$ ($y\to 1$), equation~(\ref{cuspNFW}) becomes $\rho(r)/\rhos=x^{-1.5}(1+x)^{-1.5}$, which transitions from $\rho\propto r^{-1.5}$ when $r\ll\rs$ to $\rho\propto r^{-3}$ when $r\gg\rs$ and has been noted to be a good description of prompt cusps without significant further accretion \citep{2018PhRvD..97d1303D,2018PhRvD..98f3527D}. Note that equation~(\ref{cuspNFW}) only makes sense for $A\leq\rhos\rs^{1.5}$ (or $y\leq 1$). Physically this is a lower limit on $\rhos\rs^{1.5}$, not an upper limit on $A$ (which is set directly by the initial conditions).\footnote{In terms of the concentration parameter $c$ in equation~(\ref{c}), the $y\leq 1$ restriction translates to the condition $200\bar\rho\,M_{200}/A^2\geq 384\pi[\arsinh\sqrt{c/2}-\sqrt{c/(2 + c)}]^2/c^3$. Note that the right-hand side monotonically decreases with increasing $c$, so this expression may be interpreted as a lower limit on $c$ for fixed $M_{200}$ and $A$.}

For the general case with $0<A<\rhos\rs^{1.5}$, the cusp-NFW profile in equation~(\ref{cuspNFW}) transitions from $\rho(r)=Ar^{-1.5}$ at the smallest radii, where the prompt cusp dominates, into an NFW-like profile at larger radii. Figure~\ref{fig:profile} shows that the resulting density profile is a good description of the density profiles of the simulated halos. For the three halos in the lower panels, the black dotted curves show the best-fit cusp-NFW profile, where the fit is constrained to have the same total virial mass $M_{200}$ as the simulated halo.
For all six halos, the blue dotted curves show \textit{predicted} cusp-NFW profiles, based on the model of \citet{2013MNRAS.432.1103L}, as we discuss shortly.

Before continuing, we point out that, for the cusp-NFW profile, the mass enclosed within a radius $r$ may be written as
\begin{align}\label{cuspNFWmass}
	M(r)
	&=
	4\pi\rs^3\rhos
	\Bigg[
	2\arsinh\frac{\sqrt{x}}{y}-\frac{\sqrt{x(x+y^2)}}{1+x}
	\nonumber\\&\hphantom{=4\pi\rs^3\rhos\bigg[}
	-\frac{2-y^2}{\sqrt{1-y^2}}\artanh\sqrt{\frac{x(1-y^2)}{(x+y^2)}}
	\,\,\Bigg].
\end{align}
Here $x = r/\rs$ and $y = A/(\rhos \rs^{1.5})$ as before, and $\arsinh$ and $\artanh$ are the inverse hyperbolic sine and tangent functions, respectively.
Additionally, we show in appendix~\ref{sec:stability} that a halo with a cusp-NFW profile is stable.

\subsection{Connection to accretion history}\label{sec:cuspNFWacc}

As \citet{2023MNRAS.518.3509D} emphasized, a prompt cusp is created by the collapse of the initial density peak, whereas the broader halo density profile is set by the subsequent mass accretion history.
Motivated by this consideration, we now attempt to build the broader density profile using the results of \citet{2013MNRAS.432.1103L}, which quantified how halo density profiles are related to their mass accretion histories.

The analysis in \citet{2013MNRAS.432.1103L} centers around predicting the radius $r_{-2}$ at the logarithmic slope of the density profile is $\diff\ln\rho/\diff\ln r=-2$. For the NFW profile this would simply be $r_{-2}=\rs$, but for the cusp-NFW profile of equation~(\ref{cuspNFW}), one can show that
\begin{align}\label{r2}
	r_{-2} &= \left(1 - \frac{3}{2}y^2 + \sqrt{1-y^2+\frac{9}{4}y^4}\right) \frac{\rs}{2}.
\end{align}
Following standard convention, we define the halo concentration parameter as
\begin{align}\label{c}
	c &= R_{200}/r_{-2},
\end{align}
where $R_{200}=[3M_{200}/(4\pi\, 200\bar\rho)]^{1/3}$ is the halo virial radius.
Now consider the mass $M_{-2}$ enclosed within $r_{-2}$.
\citet{2013MNRAS.432.1103L} found that the average density inside $r_{-2}$ is proportional to the density of the universe when the halo had mass $M_{-2}$, i.e.,
\begin{equation}\label{L13}
	\frac{M_{-2}}{(4\pi/3) r_{-2}^3}\simeq \omega \bar\rho|_{M_{200}=M_{-2}},
\end{equation}
with proportionality factor $\omega\simeq 776$.

Equation~(\ref{L13}) connects the concentration parameter $c$ to the past mass accretion history in the following way. Let $M=M_{200}$ be the halo mass at the ``present'' time when we wish to evaluate $c$, and let $\bar\rho$ be the mean density of the universe at the same time. At each previous time when the density of the universe is $\bar\rho'>\bar\rho$, the halo mass is $M'<M$. The narrow panels in figure~\ref{fig:profile} show the accretion history of each halo in this $(\bar\rho'/\bar\rho,M'/M)$ form.
But for an NFW profile, equation~(\ref{L13}) implies that the halo concentration $c$ satisfies
\begin{align}\label{L13a}
	\left(\frac{\bar\rho'}{\bar\rho},\frac{M'}{M}\right)
	=
	\frac{[\ln(4)-1]/2}{\ln(1 + c)-c/(1 + c)}
	\left(\frac{200}{\omega} c^3,1\right)
\end{align}
for some point $(\bar\rho'/\bar\rho,M'/M)$ on the halo's mass accretion history.
The same relationship is approximately valid for the cusp-NFW profile, since it differs only marginally from NFW at the relevant radii $r\gtrsim r_{-2}$.
The blue curves in the same panels of figure~\ref{fig:profile} show this parametric relationship, with the $c$ values labeled.
All that remains is to infer the concentration parameters by finding where equation~(\ref{L13a}) intersects each halo's mass accretion history.

The blue dotted curves in figure~\ref{fig:profile} show the cusp-NFW profiles with the concentration parameters predicted through this procedure. For the halos in the upper panels, these predictions match the simulated halos remarkably well. We emphasize that we did not even need to perform a fit. By using the cusp-peak connection in equation~(\ref{cusp}) to predict $A$ and the results of \citet{2013MNRAS.432.1103L} to predict the halo concentration $c$, and hence $\rhos$ and $\rs$, we obtain a cusp-NFW profile that accurately matches the simulation result.

For the lower panels, the predicted cusp-NFW profile does not match the simulation so precisely. For these cases, we separately identify the best-fitting concentration parameter $c$, and it is about 20 to 30 percent smaller than the concentration parameter predicted from the results of \citet{2013MNRAS.432.1103L}. The blue dotted curves show the predicted density profiles, while the black dotted curves show the fits.

The main difference between halos in the upper and lower panels of figure~\ref{fig:profile} is the value of $y=A/(\rhos\rs^{1.5})$.
For halos C1, C3, and W3 in the upper panels of figure~\ref{fig:profile}, the predicted profiles have $y\simeq 1/11$, $1/7$, and $1/6$, respectively, and worked well.
For halos W5, H1, and H3 in the lower panels, the predicted profiles have $y\sim 0.3$ and the best-fitting profiles have $y\sim 0.4$. These larger $y$ indicate that the central regions ($r\lesssim r_{-2}$) are more dominated by the initial prompt cusp and less influenced by later accretion.
Apparently, for such halos, the procedure of \citet{2013MNRAS.432.1103L} tends to somewhat overestimate halo concentrations. Indeed this finding is not new; \citet{2016MNRAS.460.1214L} found that a modified approach to accretion histories is needed to accurately describe halos close to the cutoff scale, in which they consider the summed accretion history of all progenitor halos (above some threshold) instead of only the main progenitor.

Overall, we conclude from this exercise that previous models of halo concentration parameters are suitable to use with the cusp-NFW profile.
The results of \citet{2013MNRAS.432.1103L} produce accurate density profiles in most regimes, only becoming mildly inaccurate for halos that are significantly influenced at their $r_{-2}$ radius by the cutoff in the power spectrum. Moreover, for such halos, the inaccuracy was already known and accounted for by other works, such as \citet{2016MNRAS.460.1214L}.\footnote{\citet{2016MNRAS.460.1214L} find a version of equation~(\ref{L13}) with $\omega\simeq 400$ (instead of 776), but where the mass accretion history is now taken to include the total mass in progenitor halos more massive than $1/50$ the final halo mass. If we neglect the latter consideration and simply use $\omega=400$ with the main-progenitor accretion histories in figure~\ref{fig:profile}, we obtain predictions of $c=3.3$ for halo W5, $c=4.5$ for halo H1, and $c=24$ for halo H3. These may be viewed as lower-limit predictions, since incorporating additional progenitors in the mass accretion history would yield larger $c$. Promisingly, these predictions lie below (and sometimes very close to) the best-fitting $c$ in figure~\ref{fig:profile}.}

Thus, for given halo mass $M$, the cusp coefficient $A$ follows from the cusp-halo relation in section~\ref{sec:cusphalo}, while the concentration parameter $c$ may be estimated using previous concentration-mass relations developed for halos close to the cutoff scale in the power spectrum, such as that developed by \citet{2016MNRAS.460.1214L} or the simpler fitting forms of \citet{2016MNRAS.455..318B} or \citet{2020Natur.585...39W}. The parameters $M$, $A$, and $c$ then fully determine the scale parameters $\rhos$ and $\rs$ of the cusp-NFW density profile in equation~(\ref{cuspNFW}). The scale radius $\rs$ follows by inverting equation~(\ref{r2}) with $r_{-2}=R/c$, where $R$ is the virial radius determined from the mass $M$. Explicitly, the inverse of equation~(\ref{r2}) is
\begin{align}
	\rs &= \left[\left(\frac{1}{3}-\frac{\gamma^2}{2}\right)\frac{1}{\zeta}+\frac{1+\zeta}{3}\right]
	r_{-2},
	\qquad
	\gamma = \frac{A}{\rhos r_{-2}^{1.5}},
	\nonumber\\
	\zeta &= \left(1 + 18\gamma^2 + \frac{3}{4}\gamma\sqrt{72 + 564\gamma^2 + 6\gamma^4}\right)^{1/3}.
\end{align}
Then $\rhos$ is determined by demanding that the total mass $M(c r_{-2})$, with $M(r)$ given by equation~(\ref{cuspNFWmass}), match the virial mass $M$. This match must be made numerically. For convenience, we include algorithms implementing the $(M,c,A)\to(\rs,\rhos,A)$ transformation in the provided $\textsc{Python}$ package.

As a very general but approximate alternative to previous halo concentration models, we also present in appendix~\ref{sec:concentration} an estimate of halo concentrations that is based on the mass accretion history suggested by the cusp-halo relation in section~\ref{sec:cusphalo}. Since this mass accretion history is approximately universal, this estimate produces a single characteristic concentration value at each epoch for all halo masses.
Remarkably this estimate nevertheless predicts the median concentration to within about 30 percent over a wide range of halo masses, both in our simulations and in the warm dark matter simulations of \citet{2016MNRAS.455..318B}. We include algorithms implementing this halo concentration estimate in the provided \textsc{Python} package.

\subsection{Examples}

\begin{figure}
\centering
\includegraphics[width=\columnwidth]{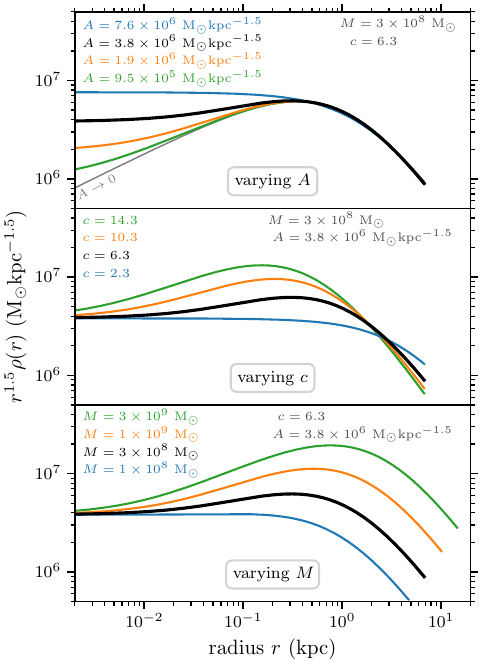}
\caption{Examples of cusp-NFW density profiles (equation~\ref{cuspNFW}) for halos of warm dark matter, plotted in the scaled form $r^{1.5}\rho(r)$. The black curve in each panel shows the typical density profile for a halo of mass $M=3\times 10^{8}~\Msol$ at redshift $z=2$ for dark matter of particle mass $m_\chi=10$~keV; here the cusp coefficient $A=3.8\times 10^6~\Msol\kpc^{-1.5}$ is predicted using the cusp-halo relation (section~\ref{sec:cusphalo}) while the concentration $c=6.3$ is predicted in accordance with appendix~\ref{sec:concentration}. In the upper panel we show profiles of different $A$, in the middle panel we vary $c$ instead, and in the lower panel we vary the halo mass $M$.}
\label{fig:wdm_prof}
\end{figure}

Figure~\ref{fig:wdm_prof} shows examples of cusp-NFW density profiles with different parameters. For the black curve, we choose a halo of mass $3\times 10^8~\Msol$ and adopt a warm dark matter cosmology with particle mass 10~keV. Then we select at redshift $z=2$ the predicted cusp coefficient $A$ (from section~\ref{sec:cusphalo}) and concentration $c$ (from appendix~\ref{sec:concentration}) for such a halo.
In the upper panel, the light gray curve shows the NFW profile with the same concentration. The presence of the central prompt cusp begins to significantly alter the structure of the halo below a radius of about 100~pc.

In the upper panel of figure~\ref{fig:wdm_prof}, we vary the cusp coefficient $A$ at fixed halo mass and concentration. For example, this change could correspond to the effect of different dark matter particle masses on the same halo. In the middle panel, we vary the concentration parameter $c$ while keeping the cusp and halo mass fixed. Finally, in the lower panel, we show the effect of varying the halo mass $M$ while keeping the cusp coefficient and halo concentration fixed.

\section{Conclusion}\label{sec:conclusion}

Every halo of collisionless particle dark matter has a $\rho=A r^{-1.5}$ density cusp at its center, which is a relic of its initial condensation out of the smooth density field of the early Universe. In this work, we developed a \textit{cusp-halo relation} that prescribes how to evaluate the coefficient $A$ of each halo's central cusp. We also proposed a ``cusp-NFW'' density profile that provides a unified description of a halo with a prompt cusp at its center.
To enable easy access to the results of this work, we provide a \textsc{Python} package at \url{https://github.com/delos/cusp-halo-relation} \citep[][]{delos_2025_17064050} that implements the cusp-halo relation and the cusp-NFW profile.

To develop the cusp-halo relation, we made use of the simulations of \citet{2023MNRAS.518.3509D}, which considered a range of idealized cosmologies.
We found that the cusp-halo relation is nearly universal across these cosmologies. However it is not exactly universal, and an avenue for future work will be to evaluate the parameters of the cusp-halo relation for specific cosmologies of interest, such as warm or interacting dark matter models.
As we illustrated in figure~\ref{fig:diagram}, the cusp-halo relation is constructed by tracing a halo's mass accretion history backwards in time until it meets the (time-dependent) distribution of newly formed, ``young'' prompt cusps. There are thus two main ingredients: the mass accretion history and the properties of young cusps. Each of these ingredients may be straightforwardly updated to match future simulation results.

Our simulation analysis took advantage of previous results that found a tight connection between prompt cusps and the initial density peaks that form them \citep{2019PhRvD.100b3523D,2023MNRAS.518.3509D,2024MNRAS.52710802O}. Instead of attempting to spatially resolve each cusp, we identified all density peaks in the initial conditions and used the simulation particles originating from each peak as tracers of the expected prompt cusp.
This choice means that the accuracy of our results depends on the accuracy to which the peak-cusp connection was calibrated in previous works. However, incorporating any future update to these calibrations into the cusp-halo relation is straightforward.
This procedure also means that we assume cusp properties do not change during halo growth and clustering, but we showed that halo mergers are only expected to cause cusp coefficients $A$ to change by around 10 percent.

Like most studies of halo density profiles, the results of this work are directly applicable to field halos. To apply these results to dark matter subhalos, the cusp-halo relation and the cusp-NFW profile must be interpreted as descriptions of the pre-infall state of the subhalo. Subhalo evolution due to tidal forces from a host halo may then be accounted for using existing models of tidal evolution \citep[e.g.][]{2020MNRAS.494..378D,2020MNRAS.498.3902Y,2022MNRAS.516..106D,2022MNRAS.517.1398B,2023MNRAS.521.4432S,2024ApJ...971...91R,2024PhRvD.110b3019D}. To assist with dynamical modeling, the provided \textsc{Python} package includes algorithms to evaluate for the cusp-NFW profile such quantities as the enclosed mass profile, the gravitational potential, the velocity dispersion, and the distribution function.

Also, this work applies most directly to ``collisionless'' dark matter with no significant nongravitational interactions. Self-interacting and baryon-interacting dark matter models have been of interest in recent years \citep{2018PhR...761....1B,2018PhR...730....1T,2019A&ARv..27....2S,2022PhR...961....1B}. Moreover, these models can naturally produce a cutoff in the spectrum of density perturbations in the early Universe \citep[e.g.][]{2009NJPh...11j5027B,2014PhRvD..89b3519D,2016PhRvD..93l3527C,2016MNRAS.460.1399V,2018PhLB..783...76H,2021JCAP...05..013E,2022JCAP...07..012G}, which would lead to prompt cusp formation.
For these models, the results of this work should be taken as the initial condition for a halo's collisional evolution.

Prompt cusps will be important to account for in observational studies of warm dark matter. Previous studies focused on how warm dark matter suppresses the abundance and internal density of low-mass halos.
However, prompt cusps give rise to an opposing effect, making low-mass subhalos more centrally compact.
Properly accounting for these features could modify observationally inferred limits on the properties of the dark matter particle. Prompt cusps could also give rise to new, independent opportunities to probe the nature of dark matter \citep[e.g.][]{2023MNRAS.522L..78D}.
This work facilitates these programs by providing a simple framework for accurately including prompt cusps in a halo or subhalo model.

%
\section*{Acknowledgments}
The author thanks Andrew Benson, Daniel Gilman, Ethan Nadler, and Neal Dalal for valuable discussions. The author also thanks Andrew Benson and Simon White for helpful comments on the manuscript.




\software{\textsc{NumPy} \citep{harris2020array}, \textsc{SciPy} \citep{2020SciPy-NMeth}, \textsc{Matplotlib} \citep{Hunter:2007}.}


\appendix

\section{Growth histories and halo concentrations}\label{sec:concentration}

We derived the cusp-halo relation by assuming in equation~(\ref{Mgrowth}) that every halo grew in mass from its initial cusp state in accordance with the same time-dependent growth factor $\chi(\sigma_0)$ given by equation~(\ref{massgrowth}), irrespective of halo mass and even of cosmology.
In principle, the cusp-halo relation only requires that this relationship hold at the cusp formation time and the ``final'' time (at which we are analyzing the halo), and not necessarily at intermediate times.
Nevertheless, this outcome suggests that there may be an approximately universal growth history for halos close to the cutoff scale in the power spectrum. Here we test this hypothesis and examine its implications for halo structures.

\subsection{A universal mass accretion history?}

Equations (\ref{massgrowth}) and~(\ref{Mgrowth}) suggest a universal mass accretion history of the following form. For a halo of mass $M$ at time $\sigma_0$, the mass $M'$ of its main progenitor at an earlier time $\sigma_0'<\sigma_0$ is
\begin{align}\label{mah}
	M' \simeq \e^{\kappa(1/\sigma_0-1/\sigma_0')}M
\end{align}
with $\kappa\simeq 4.5$ in accordance with equation~(\ref{massgrowth}).
The functional form of this mass accretion history matches the $M'/M=\e^{-\alpha_\mathrm{MAH} z}$ form suggested by \citet{2002ApJ...568...52W}, where $z$ is redshift and $\alpha_\mathrm{MAH}$ is a fitting parameter.
However, we are not aware of any previous suggestion that this mass accretion history should be applicable to halos close to the cutoff scale in the power spectrum, let alone that for these halos the fitting parameter $\kappa$ should be approximately independent of halo mass and cosmology.

\begin{figure}
	\centering
	\includegraphics[width=\columnwidth]{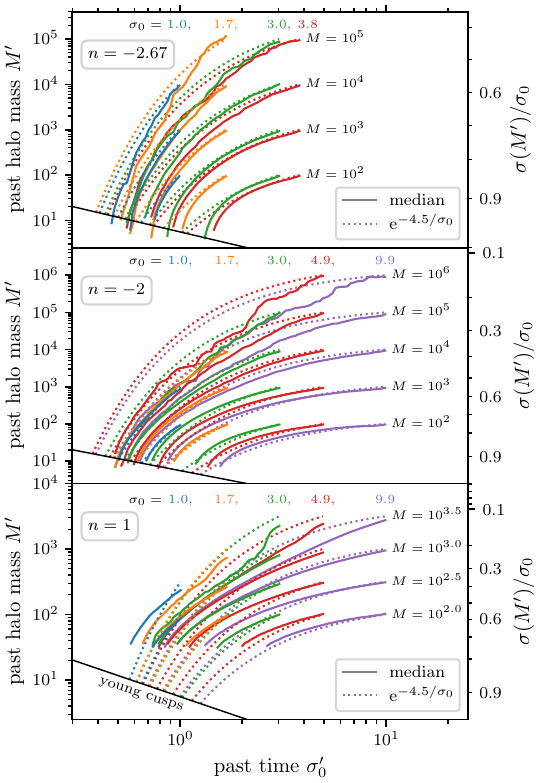}
	\caption{Mass accretion histories of the halos in our simulations (different panels). We consider halos at several different times $\sigma_0$, labeled at the top of each panel, and in several different mass ranges, labeled on the right. In each case we select halos in the mass range $(10^{-1/4}M,10^{1/4}M)$ at $\sigma_0$, where $M$ is the labeled mass, and show the median mass at previous times $\sigma_0'$. We only consider masses larger than $32m_\mathrm{p}$, where $m_\mathrm{p}$ is the simulation particle mass.
	The dotted curves show the mass accretion histories that correspond to equation~(\ref{mah}); they match the simulated mass accretion histories reasonably well at sufficiently low halo masses.
	On the right, we show the rms density contrast (in linear theory) on each mass scale.
	The black diagonal line marks the characteristic mass of newly formed cusps at each time (equation~\ref{mAchar}). See table~\ref{tab:units} for units.
	}
	\label{fig:massgrowth}
\end{figure}

Figure~\ref{fig:massgrowth} shows the mass accretion histories in our simulations. For a range of ``final'' times, we select halos by their mass at that time and show their past mass growth. The dotted curves show equation~(\ref{mah}) for comparison. Although the mass accretion history in equation~(\ref{mah}) is not accurate in all regimes, it is reasonably accurate for $M\lesssim 10^4$ in the $n=-2.67$ and $n=-2$ simulations and $M\lesssim 10^3$ in the $n=1$ simulation. For reference, we also show (on the right) the rms density contrast on each mass scale, defined as
\begin{equation}
	\sigma^2(M)=\int_0^\infty\frac{\diff k}{k}\mathcal{P}(k)W^2(k R),
\end{equation}
where $R=[3M/(4\pi\bar\rho)]^{1/3}$ is the radius of the sphere of mass $M$ and $W(x)=(3/x^3)(\sin x-x\cos x)$ is the Fourier transform of the top-hat window function. Note that $\sigma(M)<\sigma_0$ for all positive $M$ (cf. equation~\ref{sigmaj}).

\begin{figure}
	\centering
	\includegraphics[width=\columnwidth]{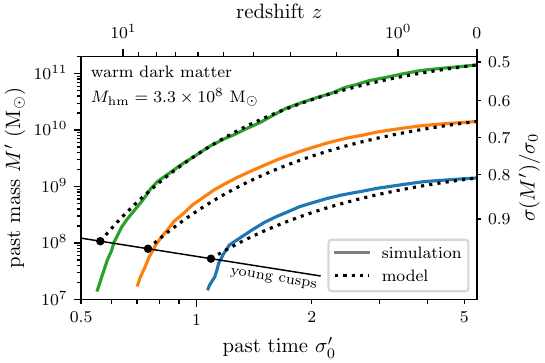}
	\caption{Mass accretion histories of halos of mass $M=1.4\times 10^{9}~\Msol$ (blue), $M=1.4\times 10^{10}~\Msol$ (orange), and $M=1.4\times 10^{11}~\Msol$ (green) at redshift 0 in the warm dark matter simulation of \citet{2016MNRAS.455..318B}, for which the half-mode mass is $3.3\times 10^8~\Msol$. The dotted curves show for each halo mass the accretion history corresponding to equation~(\ref{mah}), which matches reasonably well. The solid black line shows the characteristic mass of newly formed prompt cusps in each epoch, according to equation~(\ref{mAchar}).
	These are shown as a function of the rms (linear-theory) density contrast, which grows in time. The corresponding redshift is shown at top.
	On the right, we show for reference the rms (linear-theory) density contrast on each mass scale.
	}
	\label{fig:wdm_m}
\end{figure}

As an independent test, the colored lines in figure~\ref{fig:wdm_m} show the main-progenitor mass accretion histories \citep[presented by][]{2016MNRAS.460.1214L} that arise in the warm dark matter simulations of \citet{2016MNRAS.455..318B}. These simulations adopt a cosmology in line with WMAP measurements \citep{2011ApJS..192...18K}, and a cutoff is imposed on the initial matter power spectrum in accordance with the prescription of \citet{2005PhRvD..71f3534V} for a 3.3~keV dark matter particle \citep[but see][]{2023PhRvD.108d3520V}.
Using a matter power spectrum from \citet{1998ApJ...496..605E} with the same cosmological parameters, we evaluate $\sigma_0$ as a function of time for this cosmology and overplot in black the mass accretion history from equation~(\ref{mah}).
Although it exhibits some systematic offset, this mass accretion history works reasonably well for halos up to more than $10^{11}~\Msol$ at redshift 0. The mass definition in this case is $M_{200\mathrm{c}}$, the mass in the sphere that encloses average density 200 times the critical density.

\subsection{Implication for halo concentrations}

\begin{figure}
	\centering
	\includegraphics[width=\columnwidth]{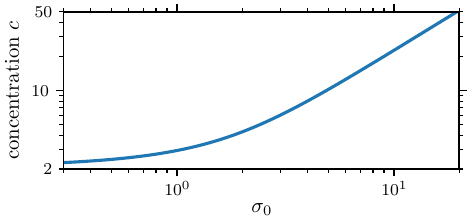}
	\caption{Time evolution of the concentration parameter $c$ of all halos based on the mass accretion history in equation~(\ref{mah}). We show how $c$ depends on the rms density contrast (in linear theory), which grows in time.}
	\label{fig:concentration-time}
\end{figure}

According to the results of \citet{2013MNRAS.432.1103L}, the mass accretion history in equation~(\ref{mah}) should imply that halo concentrations are a function of $\sigma_0$ (time) and independent of halo mass. For this mass accretion history, figure~\ref{fig:concentration-time} shows the time evolution of the concentration $c$, according to the model of \citet{2013MNRAS.432.1103L} applied as in section~\ref{sec:cuspNFWacc}. Here we assume a matter-dominated universe. Halo concentrations are $c\simeq 2.1$ when $\sigma_0\ll 1$ and grow over time, scaling slightly faster than linearly with $\sigma_0$ at late times.

\begin{figure}
	\centering
	\includegraphics[width=\columnwidth]{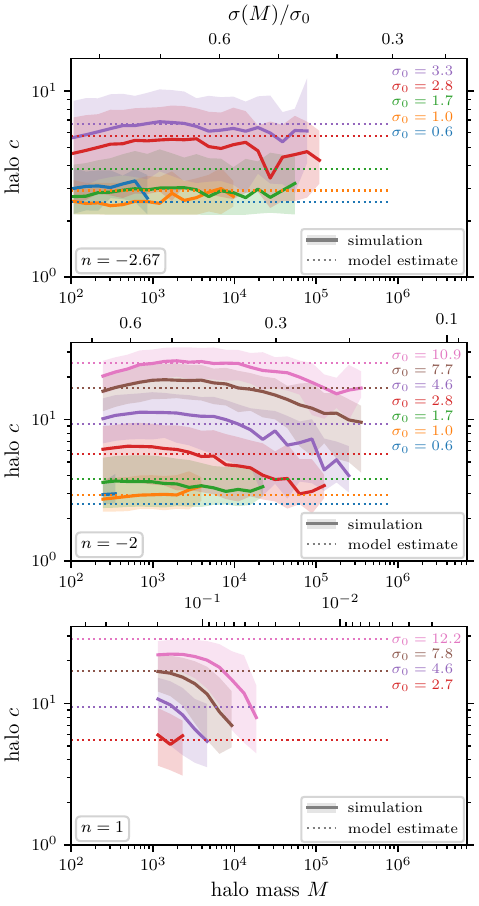}
	\caption{Concentration-mass relations in the $n=-2.67$, $n=-2$, and $n=1$ simulations (different panels). The solid lines show the median concentration at each mass (in bins of width $\Delta\ln M=0.35$); we plot this only down to a mass of $10^3m_\mathrm{p}$, where $m_\mathrm{p}$ is the simulation particle mass, due to the necessity of resolving the internal structures of halos. The shaded bands mark the 16th and 84th percentiles. Different colors correspond to different times. For comparison, the dotted lines indicate the concentration parameter predicted at each time from the mass accretion history in equation~(\ref{mah}). At sufficiently low masses, this simple estimate yields halo concentrations that are accurate at all times to within about 30 percent. See table~\ref{tab:units} for units.}
	\label{fig:concentration}
\end{figure}

We now test this prediction in our simulations. For each field halo, we evaluate the radius $r_\mathrm{max}$ at which the circular orbit velocity (or equivalently $M(r)/r$, where $M(r)$ is the mass enclosed in radius $r$) is maximized. Now we estimate the radius $r_{-2}$ as $r_{-2}\simeq r_\mathrm{max}/2.2$, which is valid for NFW-like density profiles. The concentration is $c=R_{200}/r_{-2}$ as in equation~(\ref{c}).
Figure~\ref{fig:concentration} shows the resulting concentration distributions as a function of halo mass for each of the three simulations.
Although concentrations are not independent of mass, they are nearly so for halos close in mass to the cutoff scale in the power spectrum.
Moreover, for these halos, the predicted concentration is accurate to within about 30 percent at all times.

\begin{figure}
	\centering
	\includegraphics[width=\columnwidth]{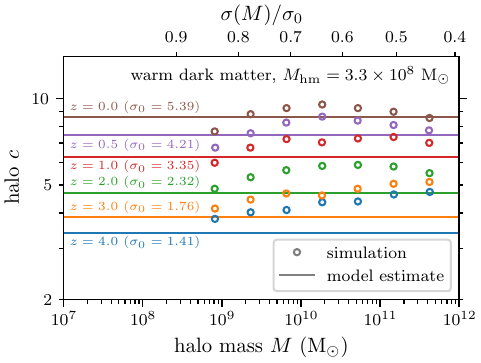}
	\caption{Concentration-mass relation in the warm dark matter simulation of \citet{2016MNRAS.455..318B}, for which the half-mode mass is $3.3\times 10^8~\Msol$. The colored circles show median concentrations over a range of times (different colors). The lines mark the concentration that would be predicted at each time from the mass accretion history in equation~(\ref{mah}). It is accurate to within about 30 percent.}
	\label{fig:wdm_c}
\end{figure}

Figure~\ref{fig:wdm_c} shows the same test for the concentration-mass relation in the warm dark matter simulations of \citet{2016MNRAS.455..318B}. Here, too, the simple concentration estimate arising from the mass accretion history in equation~(\ref{mah}) is accurate to about 30 percent at all times.

\subsection{Scatter in halo concentrations}

At fixed halo mass, previous works \citep[e.g.][]{2002ApJ...568...52W,2015ApJ...799..108D} have noted that the 68-percent scatter in the halo concentration parameter is around 0.15 dex (corresponding to a factor of about 1.4) for halos of arbitrarily cold dark matter.
The scatter in concentration that we find in figure~\ref{fig:concentration} (shaded bands) is comparable, although it can vary considerably depending on the epoch and the halo mass bin.
Coincidentally, this level of scatter in $c$ is similar to the scatter in cusp coefficients $A$ that we found in section~\ref{sec:cusphaloscatter}.

\begin{figure}
	\centering
	\includegraphics[width=\columnwidth]{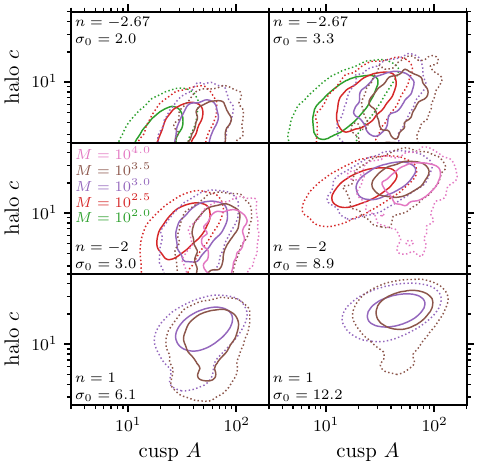}
	\caption{Combined distribution of concentration $c$ and cusp coefficient $A$ for halos in different mass bins (colors). We use the same mass bins as figure~\ref{fig:cusps_A-m} and show results from the three simulations (rows) each at two different times (columns). Solid contours enclose 68 percent of the distribution while dotted contours enclose 95 percent. $A$ and $c$ are correlated at fixed halo mass, but only weakly.}
	\label{fig:scatter_c}
\end{figure}

In figure~\ref{fig:scatter_c}, we show the combined distribution of $A$ and $c$ for a few times and halo mass bins. $A$ and $c$ both depend on the mass accretion history in similar ways, in that a slower-growing halo should have larger $c$ and $A$, while a faster-growing halo should have smaller values of both. Consequently, we expect some correlation between these parameters.
On the other hand, cusp coefficients $A$ are sensitive to the full accretion history all the way back to the halo's initial formation event, whereas halo concentrations depend only on the relatively recent accretion history, corresponding typically to a factor of 2 to 10 in halo mass (see the narrow panels in figure~\ref{fig:profile}, in which the concentration is set by the intersection of the two curves).
Consequently, the correlation between $A$ and $c$ may not be strong. Indeed, figure~\ref{fig:scatter_c} shows that these parameters are only weakly correlated at fixed time and halo mass.

\section{Stability of the cusp-NFW profile}\label{sec:stability}

\begin{figure}
	\centering
	\includegraphics[width=\columnwidth]{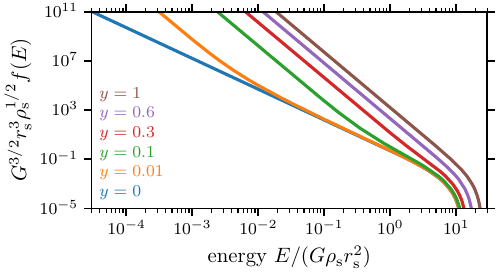}
	\caption{Distribution functions $f(E)$ that describe systems with isotropic velocity distributions and cusp-NFW density profiles of different $y=A/(\rhos\rs^{1.5})$ (different colors). We set the zero point of energy to match the gravitational potential at the center of each system.}
	\label{fig:stability}
\end{figure}

The cusp-NFW density profile places a relatively compact central mass, the prompt cusp, at the center of a broader halo with a more shallowly varying density profile. Qualitatively similar, albeit more extreme, configurations have been noted to be unstable. For example, a compact massive system near the center of a cored dark matter halo (i.e., a halo with near-uniform density at the center) is spontaneously expelled outward in a process known as dynamical buoyancy \citep{2012MNRAS.426..601C,2021ApJ...912...43B,2022ApJ...926..215B}. Therefore, we show in this appendix that a halo with a cusp-NFW profile is stable. For simplicity, we specialize to models with isotropic velocity distributions. \citet{2023MNRAS.518.3509D} found that the velocity dispersion is nearly isotropic inside a prompt cusp but can be radially anisotropic in the surrounding halo.

For halos with a range of different cusp-NFW density profiles, figure~\ref{fig:stability} shows the distribution functions $f(E)$ that represent the mass density of the system per volume in 6-dimensional position-velocity phase space. Since we assume isotropic velocity distributions, the distribution function depends only on energy $E$. We construct $f(E)$ numerically by means of the widely used inversion method of \citet{1916MNRAS..76..572E}.
The key observation is that each $f(E)$ is monotonically decreasing as a function of energy $E$. According to the combination of what \citet{1988gady.book.....B} call \textit{Antonov's second law} and the \textit{Doremus--Feix--Baumann theorem}, $f'(E)<0$ is a sufficient condition for a spherical system with an isotropic velocity distribution to be stable with respect to both radial and non-radial perturbations \citep{1961SvA.....4..859A,1965ApJ...142..229L,1971PhRvL..26..725D,1985ApJ...298...27K,1992MNRAS.259...95A}.


\newpage

\bibliography{main}{}
\bibliographystyle{aasjournal}


\end{document}